\documentclass[12pt]{article}

\newcommand{\Dal}{\drawsquare{7}{0.6}}

\def\O{{\cal O}}

\def\li2{{\rm Li}_2}

\def\det{{\rm det}}

\usepackage{paper2e}
\usepackage{mydefs2e}
\usepackage{epsf}

\def\caption#1{{\centerline{\vbox{\baselineskip=14pt
            \vskip.15in\hsize=5.5in\noindent{#1}\vskip.1in }}}}

\begin{document}

\begin{titlepage}

\preprint{UMD-PP-99-119\\
SNS-PH/99-12}

\vskip-0.2in
\title{ Soft Supersymmetry Breaking in\\\medskip
Deformed Moduli Spaces, Conformal Theories\\\medskip
and ${\cal N} = 2$ Yang-Mills Theory}


\author{Markus A.~Luty%
\footnote{Sloan Fellow.}}
\vskip .1in

\address{Department of Physics, University of Maryland\\
College Park, Maryland 20742, USA\\
{\tt mluty@physics.umd.edu}}

\author{Riccardo Rattazzi}
\vskip.1in

\address{INFN and Scuola Normale Superiore\\
I-56100 Pisa, Italy\\
{\tt rattazzi@cibs.sns.it}}

\begin{abstract}
We give a self-contained discussion of recent progress in computing
the non-perturbative effects of small non-holomorphic soft supersymmetry
breaking, including a simple new derivation of these results
based on an anomaly-free gauged $U(1)_R$ background.
We apply these results to $\scr{N} = 1$ theories with
deformed moduli spaces and conformal fixed points.
In an $SU(2)$ theory with a deformed moduli space, we completely
determine the vacuum expectation values and induced soft masses.
We then consider the most general soft breaking of
supersymmetry in $\scr{N} = 2$ $SU(2)$ super-Yang--Mills
theory.
An $\scr{N} = 2$ superfield spurion analysis is used to give an
elementary derivation of the relation between the modulus and the
prepotential in the effective theory.
This analysis also allows us to determine the non-perturbative
effects of all soft terms except a non-holomorphic scalar mass,
away from the monopole points.
We then use an $\scr{N} = 1$ spurion analysis to determine the
effects of the most general soft breaking, and also analyze the
monopole points.
We show that \naive dimensional analysis works perfectly.
Also, a soft mass for the scalar in this theory forces
the theory into a free Coulomb phase.
\end{abstract}


\end{titlepage}

\section{Introduction}
In the last several years there has been significant progress in
understanding the low-energy dynamics of strongly-coupled
supersymmetric gauge theories \cite{seiberg,seibergdeform,SW}.
Most of this progress has been limited to holomorphic quantities,
which give a great deal of interesting information if supersymmetry
(SUSY) is exact.
In many cases, the moduli space of vacua and the phase structure and
massless excitations of the theory can be exactly determined.
A natural question to ask is whether these results can be extended
to the case of explicit breaking of SUSY.
As a first step, one can study the case where SUSY is broken softly
by mass parameters that are small compared to the scale of strong
dynamics in the gauge theory.
In cases where the low-energy effective field theory is known in the
SUSY limit, one can carry out an analog of chiral perturbation theory
for SUSY breaking.

The most general soft SUSY breaking can be parameterized by turning
on higher $\th$-dependent terms in the coupling constants viewed
as superfield spurions \cite{GG}.
For example, if we write
\beq
\scr{L} = \myint d^4\th\, \scr{Z} Q^\dagger e^V Q
+ \left( \myint d^2\th\, S \tr(W^\al W_\al) + \hc \right)
+ \cdots
\eeq
with
\beq
S = \frac{1}{2 g^2} + \th^2 \frac{m_\la}{g^2},
\qquad
\scr{Z} = Z \left[ 1 - \th^2 \bar{\th}^2 m^2 \right].
\eeq
then $m_\la$ is a gaugino mass and $m^2$ is a scalar mass.
The effects of soft SUSY breaking that can be parameterized by
chiral superfields can be studied using holomorphy and SUSY
non-renormalization theorems \cite{MV,holbreak,ADKM}.
However, when studying the non-perturbative effects of soft SUSY
breaking, non-holomorphic scalar masses cannot be neglected compared
to holomorphic soft terms such as gaugino masses.
(For example, in an asymptotically free theory, if scalar masses are
smaller than gaugino masses at a renormalization scale where the
theory is weakly coupled, then the renormalization group will
generate a scalar mass comparable to the gaugino mass at the scale
where the theory becomes strongly coupled.)
In superfield language, the problem is therefore to determine how
the superfield $\scr{Z}$ in the fundamental theory couples to fields
in the low-energy theory.
\Ref{AR} pointed out that one obtains nontrivial information by
viewing $\scr{Z}$ as a gauge superfield.
The point is that $\scr{Z}$ contains a vector field
\beq
\scr{Z} = \cdots + \th\si^\mu\bar{\th} A_\mu + \cdots
\eeq
that couples to the Noether current associated with
a $U(1)$ `$Q$ number' symmetry.
As is well-known, this means that the dependence on $A_\mu$
at low energies are controlled simply by charge conservation.
SUSY relates this to the dependence on the soft mass, and one
obtains non-perturbative information about
non-holomorphic SUSY breaking at low energies.

To make this idea precise, one must deal with several technical
complications.
First, one must understand the renormalization properties of the
superfield couplings \cite{NSVZ,HS,AGLR}.
Second, $U(1)$ `gauge' symmetries such as the one discussed above are
generally anomalous.
This does not give rise to any inconsistency (the relevant
gauge fields are non-dynamical sources),
but it does mean that the $U(1)$
symmetry is broken explicitly, and this must be properly taken into
account.
These problems were addressed in \Ref{AR} in a `Wilsonian' language,
and used to obtain results in several theories of interest.

In the present paper, we extend the results of \Ref{AR} in several
ways.
First, we give a self-contained review of the method of \Ref{AR}
in terms of renormalized couplings and superfield RG invariants.
We also give a new derivation based on a non-anomalous gauged
$U(1)_R$ symmetry in a supergravity background.
We apply these results to several classes of $\scr{N} = 1$ theories
that were not treated in \Ref{AR}, namely those with deformed moduli
spaces and conformal fixed points.
In the $SU(2)$ theory with a deformed moduli space, we are able to
determine the vacuum uniquely for vanishing gaugino masses,
and compute the soft masses of the composite fields for arbitrary
perturbations at the maximally symmetric point.
In the conformal window of SUSY QCD, we give a very simple
derivation of the fact that soft masses scale to zero as one
approaches the fixed point.
This result was previously obtained in \Ref{KKKZ} by
explicit calculation.

We then turn to $\scr{N} = 2$ $SU(2)$ super Yang--Mills theory.
This was studied in \Ref{ADKM} for a special subset of the possible
soft SUSY breaking terms.
We generalize these results to include the non-perturbative effects of
the most general possible soft SUSY breaking.
We first perform a spurion analysis in terms of $\scr{N} = 2$
superfields that includes all soft breaking terms except a non-holomorphic
scalar mass.
This analysis also leads to an elementary derivation of the relation
between the modulus and the effective prepotential that had
previously been obtained using properties of the Seiberg--Witten
solution.
We then analyze the theory using the $\scr{N} = 1$ techniques discussed
above.
In this way, we are able to determine the exact potential on the
full moduli space for general soft SUSY breaking, including
the potential near the monopole points.
The agreement between the two calculations serves
as a nontrivial check on our methods.

We find a rich structure of phase
transitions in this theory as a function of the soft masses.
For example, when fermion masses dominate, the vacuum is near the
monopole/dyon points and exhibits confinement via monopole/dyon
condensation;
when the scalar mass dominates, the vacuum is at the origin and
the theory is in a Coulomb phase.
We also show that `\naive dimensional analysis' works perfectly
for all the quantities we compute, giving strong support for these
methods in strongly coupled SUSY theories.

\section{Non-perturbative Non-holomporphic ${\cal N} = 1$ Soft SUSY Breaking}
In this Section, we review the results of \Ref{AR}
on the non-perturbative effects of soft $\scr{N} = 1$
SUSY breaking, including non-holomorphic scalar masses.
Our discussion here uses renormalized couplings
rather than the `Wilsonian' language of \Ref{AR},
but all results are completely equivalent.
We then apply this formalism to the case of deformed moduli spaces.
This case has not been considered before, so the results are
interesting in their own right.
This case also has some important
similarities with the $\scr{N} = 2$ super
Yang--Mills theory we will study in the following Section.

\subsection{RG Invariant Superfield Spurions}
Consider a \susc gauge theory defined in the ultraviolet
by a renormalized lagrangian
at a scale $\mu$ where the theory is weakly
coupled.
We consider here only the case where the gauge group is simple
and there are no superpotential terms.
The renormalized lagrangian in superspace is
\beq \eql{lagrangian}
\scr{L}(\mu) = \myint d^4 \th\,
\sum_r \scr{Z}_r(\mu) \Phi_r^\dagger e^{V^{(r)}} \Phi_r
+ \left(
\myint d^2\th\, S(\mu) \tr (W^\al W_\al) + \hc
\right).
\eeq
The renormalized couplings $\scr{Z}_r(\mu)$ and $S(\mu)$ can be
promoted to superfields to all orders in perturbation theory
\cite{AGLR}, and SUSY breaking can be included by non-zero
$\th^2$ and $\bar{\th}^2$ dependence in the couplings:
\beq
\scr{Z}_r(\mu) &= Z_r(\mu) \left[
1  - \th^2 B_r(\mu) - \bar{\th}^2 B_r^\dagger(\mu)
- \th^2 \bar{\th}^2 \left( m_r^2(\mu) - |B_r(\mu)|^2 \right) \right],
\\
S(\mu) &= \frac{1}{2 g_S^2(\mu)} - \frac{i\Th}{16\pi^2}
+ \th^2 \frac{m_{\la S}(\mu)}{g_S^2(\mu)}.
\eeq
The quantity $\scr{Z}_r$ is a real superfield.
Its components are defined so that
$Z_r$ is the usual wavefunction factor, $B_r$ is a $B$-term.
An elementary but important point is that the $\th^2$ terms affect the
equation of motion for the auxiliary fields, with the result that the
physical soft mass depends on the logarithm of the superfield
$\scr{Z}_r$:
\beq
m_r^2 = -[ \ln \scr{Z}_r ]_{\th^2 \bar{\th}^2}.
\eeq

The quantity $S$ is chiral and runs only at one loop
\cite{russ}.
Its components are defined so that $\Th$ is the vacuum angle,
and at 1-loop level, $g_S$ is the gauge coupling and $m_{\la S}$
is the gaugino mass.
However, $g_S$ and $m_{\la S}$ differ from the
conventionally-defined renormalized gauge coupling $g$ and
gaugino mass $m_{\la}$ at two loops and beyond \cite{russ}.
One manifestation of this is the fact that under the transformation
\beq[onetrans]
\bal
\Phi_r &\mapsto e^{A_r} \Phi_r,
\\
\scr{Z}_r(\mu) &\mapsto e^{-(A_r + A_r^\dagger)} \scr{Z}_r(\mu),
\eal\eeq
where $A_r$ is a constant chiral superfield,
the coupling $S$ has an anomalous transformation
\beq\eql{Kanom}
S(\mu) \mapsto S(\mu) - \sum_r \frac{t_r}{8\pi^2} A_r.
\eeq
Here, $t_r$ denotes the index of the representation $r$.%
\footnote{The index $t_r$ is normalized to $\frac{1}{2}$ for fundamentals.}
For $A_r$ pure imaginary, \Eq{onetrans} is a
$\U1 \times \cdots \times \U1$ transformation with charges
\beq
\scr{Q}_r(\Phi_s) = +\de_{rs},
\eeq
and \Eq{Kanom} is a manifestation of the chiral anomaly.
For $A_r$ pure real, \Eq{onetrans} is a rescaling of the
fields under which physical quantities are invariant, and
\Eq{Kanom} is a manifestation of the Konishi (field rescaling)
anomaly \cite{Konishi}.
Note that $\scr{Z}_r$ transforms as a $U(1)$ gauge superfield.
The non-perturbative validity of these `anomalous $\U1$'
symmetries is the crucial new ingredient introduced in \Ref{AR}
to analyze the non-perturbative effects of soft masses in the theory.

The couplings $g$ and $m_\la$
are the lowest components of a \emph{real} superfield
\beq
R(\mu) &= \frac{1}{g^2(\mu)}
+ \left( \th^2 \frac{m_\la(\mu)}{g^2(\mu)} + \hc \right)
+ \cdots
\nonumber\\
&= S(\mu) + S^\dagger(\mu)
+ \frac{t_G}{8\pi^2} \ln R(\mu)
- \sum_r \frac{t_r}{8\pi^2} \ln \scr{Z}_r(\mu)
+ \scr{O}(R^{-1}).
\eeq
$R$ is invariant under the transformation \Eq{onetrans}.
One can choose a special `NSVZ' scheme in which all of the
$\scr{O}(R^{-1})$ and higher corrections vanish, but this will
not be important for our results.
For a discussion of the superfield $R$ (and in particular the
role of its $\th^2 \bar{\th}^2$ component) see
\Ref{AGLR}.

If this theory is asymptotically free, the strong
dynamics occurs at a scale $\mu \sim \La$ where the gauge coupling
becomes large.
The scale $\La$ must clearly be RG invariant.
With the ingredients above, we see that we can form \emph{two}
RG-invariant scales:
\beq\eql{scales}
\La_S \equiv \mu \exp \left\{ -\frac{16\pi^2 S(\mu)}{b} \right\},
\qquad
\La_R \equiv \mu \exp\left\{ -\int^{R(\mu)} \frac{dR}{\be(R)} \right\},
\eeq
where
\beq
b = 3 t_G - \sum_r t_r,
\qquad
\be(R) \equiv \mu \frac{d R}{d \mu}.
\eeq
The RG invariant scale
$\La_S$ is a chiral superfield, and transforms under
\Eq{onetrans} as
\beq\eql{lastrans}
\La_S \mapsto \left( \prod_r e^{2 t_r A_r / b} \right) \La_S.
\eeq
In other words, $\La_S$ is charged under the anomalous $\U1$:
\beq
\scr{Q}_r(\La_S) = \frac{2 t_r}{b}.
\eeq
Because $\La_S$ is chiral, it is the scale that appears in
non-perturbatively generated effective superpotentials.
The transformation property
\Eq{lastrans} is exactly what is required to make the effective
superpotential invariant under the anomalous $U(1)$.
For example, in a theory with a simple gauge group and vanishing
superpotential, the
\emph{anomaly-free} symmetries constrain the dynamically-generated
superpotential to have the form \cite{ADS}
\beq\eql{Weff}
W_{\rm eff} \sim \frac{1}{\La_S^{b / (t - t_G)}}
\prod_r \Phi_r^{2 t_r / (t - t_G)},
\eeq
where $t = \sum_r t_r$
is the total matter index, and the factors of $\La_S$ have been
inserted by dimensional analysis.
One can now check that the power of $\La_S$ is precisely what is
required in order for $W_{\rm eff}$ to be invariant under the
transformation \Eq{onetrans}.

The RG-invariant scale $\La_R$ in \Eq{scales}
is a real superfield
defined by analytically continuation into superspace.
Specifically, for real $R(\mu)$ \Eq{scales} defines a function of
$R(\mu)$ (up to a multiplicative constant), and the continuation into
superspace is defined by evaluating this function for $R(\mu)$ replaced
by a real superfield.
$\La_R$ is invariant under the transformation \Eq{onetrans}.

Another important RG invariant is the quantity
\beq\eql{wave}
\hat{\scr{Z}}_r \equiv
\scr{Z}_r(\mu) \exp\left\{ -\int^{R(\mu)} dR\,
\frac{\ga_r(R)}{\be(R)} \right\},
\eeq
where
\beq
\ga_r(R) \equiv \mu \frac{d \ln\scr{Z}_r}{d\mu}.
\eeq
$\hat{\scr{Z}}_r$ can be thought of as the wavefunction factor
for the field $\Phi_r$
renormalized at the RG-invariant superfield scale $\La_R$.
Like $\La_R$, the quantity
$\hat{\scr{Z}}_r$ is defined as a superfield by
analytic continuation.
Although $\hat{\scr{Z}}_r$ is RG invariant, it is not
physical by itself because it transforms under field rescalings
like a gauge superfield (like $\scr{Z}_r(\mu)$):
\beq
\hat{\scr{Z}}_r \mapsto e^{-(A_r + A_r^\dagger)} \hat{\scr{Z}}_r.
\eeq
However, $\hat{\scr{Z}}_r$ can appear in the effective lagrangian:
its transformation property is just what is required to write
kinetic terms invariant under the transformation \Eq{onetrans}.

Following \Ref{AR}, we can also define the RG-invariant
\beq\eql{hatLa}
I \equiv \La_S^\dagger \left(
\prod_r \hat{\scr{Z}}_r^{2 t_r / b} \right) \La_S,
\eeq
which is also invariant under the anomalous $\U1$'s.
It is easily shown that
\beq\eql{Ihat}
I = \hbox{constant} \times \La_R^2,
\eeq
so this does not give an independent invariant.

When the theory includes explicit SUSY breaking, the RG invariants above
have $\th$-dependent components:
\beq
\eql{large}
{}[ \ln \La_S ]_{\th^2} &=
-\frac{16 \pi^2}{b} \frac{m_{\la S}}{g_S^2},
\\
\eql{largetoo}
{}[ \ln \La_R ]_{\th^2} &=
-\frac{1}{\be(R)} [ R ]_\th^2
= -\frac{8 \pi^2}{b} \frac{m_\la}{g^2} + \cdots,
\\
{}[ \ln \hat{\scr{Z}}_r ]_{\th^2} &= -B_r
-\frac{\ga_r(R)}{\be(R)} [ R ]_{\th^2} = -B_r -
\frac{2 C_r}{b} m_\la + \cdots,
\\
{}[ \ln \La_R ]_{\th^2 \bar{\th}^2} &=
-\frac{1}{\be(R)} [ R ]_{\th^2 \bar{\th}^2}
+ \frac{\be'(R)}{\be^2(R)} [ R ]_{\th^2}
[ R ]_{\bar{\th}^2}
\nonumber\\
\eql{lastlarge}
&= -\frac{1}{b} \sum_r t_r \left(
m_r^2 - \frac{2 C_r}{b} |m_\la|^2 \right) + \cdots,
\\
{}[ \ln \hat{\scr{Z}}_r ]_{\th^2 \bar{\th}^2} &=
-m_r^2 - \frac{\ga_r(R)}{\be(R)} [ R ]_{\th^2 \bar{\th}^2}
- \frac{d}{dR} \left( \frac{\ga_r(R)}{\be(R)} \right)
{}[ R ]_{\th^2} [ R ]_{\bar{\th}^2}
\nonumber\\
&= -m_r^2 + \frac{2 C_r}{b} |m_\la|^2 + \cdots
\eeq
The ellipses denote terms that are suppressed at weak coupling;
they can be computed exactly in the NSVZ scheme, but this is not
important for our results.

The quantities in \Eqs{large}--\eq{lastlarge} are RG invariants by
construction, and can therefore be evaluated at any value of the
renormalization scale $\mu$.
In an asymptotically free theory, they simplify if
they are evaluated in the limit $\mu \to \infty$:
\beq
\eql{LaRthth}
{}[ \ln \La_R ]_{\th^2} &=
-\frac{8\pi^2}{b} \frac{m_{\la 0}}{g_0^2},
\\
{}[ \ln \hat{\scr{Z}}_r ]_{\th^2} &= -B_{r 0},
\\
\eql{LaRthththth}
{}[ \ln \La_R ]_{\th^2 \bar{\th}^2} &=
-\frac{1}{b} \sum_r t_r m_{r 0}^2,
\\
\eql{barebreakend}
{}[ \ln \hat{\scr{Z}}_r ]_{\th^2 \bar{\th}^2} &=
-m_{r 0}^2,
\eeq
where
\beq
\frac{m_{\la 0}}{g_0^2} \equiv \lim_{\mu\to\infty}
\frac{m_\la(\mu)}{g^2(\mu)},
\quad
B_{r 0} \equiv \lim_{\mu\to\infty} B_r(\mu),
\quad
m_{r 0}^2 \equiv \lim_{\mu\to\infty} m_r^2(\mu).
\eeq
We can make a field redefinition to set $B_{r 0} = 0$;
since we are considering the case where there is no superpotential,
this has no further effect.
We see that the SUSY breaking components of the RG-invariant superfield
spurions are simple combinations of the \emph{bare} coupling
constants.
This interpretation emerges very directly in the `Wilsonian'
approach of \Ref{AR}.
It is interesting and somewhat counterintuitive that the bare
scalar soft mass can be thought of as given by the wavefunction
evaluated at the scale $\mu = \La_R$
(appropriately continued into superspace).

A remark on anomaly-free generators is now in order.
The wavefunctions ${\cal Z}_r$ in \Eq{lagrangian} can be thought of
as gauge fields for the maximal abelian subgroup $[U(1)]^K$ of
the full flavor group of the model.
We can choose a basis of generators so that only one of the $U(1)$'s has
an anomaly and the rest are anomaly-free.
For a soft mass proportional to an anomaly-free generator, the RG
evolution of the soft masses is simply determined by charge conservation.
This implies that
  the mapping between the UV and IR soft masses is obtained simply by
matching quantum numbers of the composite.
For example, in SUSY QCD a soft term associated with baryon number has
the UV form $m^2_{\tilde Q}=-m^2_{\tilde {\bar Q}}=m_0^2$.
In the s-confining case $N_F = N_c + 1$,
the low-energy masses for baryons and mesons
are simply  $m^2_{\tilde B}=-m^2_{\tilde {\bar B}}=N_c m_0^2$ and
$m^2_{\tilde M}=0$.

We now consider briefly the extension of these results to theories
with superpotentials in the UV theory.
In this case, the anomalous $U(1)$ symmetries considered above do not
suffice to determine the exact dependence on the soft masses
because there are additional invariants that can be formed using
the superpotential couplings.
For example, suppose that the UV theory contains a Yukawa coupling
$\la$.
In that case, we can define an additional RG invariant $\hat{\la}$
corresponding to the running Yukawa coupling renormalized at the scale
$\La_R$.
The quantity $|\hat{\la}|^2$ is neutral under all symmetries
(including $U(1)_R$), and therefore symmetries do not suffice to
determine how this quantity appears in the effective
\Kahler potential.
We can of course use holomorphy and symmtries to determine the
exact dependence of the effective superpotential on the Yukawa coupling.
This can give nontrivial information in the case where the running
Yukawa coupling is perturbative both at the scale $\La_R$ and at a UV
scale $\mu_0$ where the gauge coupling is also perturbative.
(We cannot in general take $\mu_0 \to \infty$ because theories with
Yukawa couplings are strongly coupled in the ultraviolet.)
In that case, we can expand the RG invariants in powers of
$|\la(\mu_0)|^2 / (16\pi^2)$, and `\naive dimensional analysis'
\cite{NDA,SUSYNDA} tells us that the effective \Kahler potential is
an expansion in $|\hat{\la}|^2 / (16\pi^2)$.%
\footnote{Since $\hat{\la}$ differs from $\la(\mu_0)$ by an RG factor
of order $\la^3(\mu_0) \ln(\mu_0 / \La_R) / (16\pi^2)$, there are
large logarithms in the expansion when expressed in terms
of $\la(\mu_0)$.}
If $\la(\mu_0), \hat{\la} \ll 4\pi$, these effects are smaller than
the `tree-level' dependence on the Yukawa coupling in the effective
superpotential.

Similar remarks apply to the case where the theory has a product
gauge group, with some matter fields charged under multiple group
factors.
If one of the gauge couplings becomes strong at a scale where all
the other gauge couplings are weak, we can compute the effects of
soft masses up to perturbative corrections using ideas similar to
those discussed above for Yukawa couplings.
We cannot treat the case where several factors of the gauge group
become strong at the same scale.%
\footnote{It may be possible to make progress in theories with
discrete symmetries that interchange the gauge group factors.}

There are special choices of soft masses for which the RG-invariant
Yukawa couplings $\hat{\la}_i$ or ratios of strong scales
$\Lambda_{Ra} / \Lambda_{Rb}$ have no $\theta$ dependence.
In this case the flow of soft terms can be controlled
as in our simple SQCD examples.
The physical interpretation of these special RG trajectories is
is clarified below using a gauged $U(1)_R$ background.

\subsection{Deformed Moduli Space}
In \Ref{AR}, this formalism was applied to theories with confining
and infrared free `dual' descriptions.
We now apply these results to soft breaking in theories with deformed
moduli spaces.
We begin with $SU(2)$ SUSY QCD with
4 fundamentals $Q^j$, $j = 1, \ldots, 4$ (2 `flavors').
In the SUSY limit,
the moduli space can be parameterized by the holomorphic gauge-invariants
(`mesons')
\beq
M^{jk} = \frac{1}{\La_S^2} Q^j Q^k = -M^{kj}.
\eeq
Classically, these satisfy the constraint $\Pf(M) = 0$, but this is
modified by quantum effects to \cite{seibergdeform} (see also \Ref{MPRV})
\beq
\Pf(M) = 1.
\eeq
The anomaly-free $U(1)_R$ charge of $Q$ vanishes
and the anomalous $U(1)$ charge of $Q$ and $\La$ are the same,
so the quantum constraint is consistent with all symmetries.
To simplify the analysis, we use the (Lie algebra)
isomorphism between $SU(4)$ and $SO(6)$.
In $SO(6)$ language, we write the mesons as $M_a$, $a = 1, \ldots, 6$
with constraint
\beq\eql{soconstr}
M_a M_a = 1.
\eeq

If the soft breaking masses are small compared to the dynamical scale
$\La$ of the theory, they will make a small perturbation on the SUSY
moduli space.
We therefore write the most general $SO(6)$ invariant
effective lagrangian written in terms of fields $M$
satisfying the constraint \Eq{soconstr}.
The only $SO(6)$ invariant combinations of $M$ are
$M_a^\dagger M_a$ and $M_a M_a = 1$ (by the quantum constraint),
so we have
\beq
\scr{L}_{\rm eff} = \myint d^2\th d^2\bar{\th}\,
\La_R^2 \, k(M^\dagger M)
+ \hbox{derivative\ terms},
\eeq
where $M$ satisfies \Eq{soconstr}.
Note that $M^\dagger M$ is completely neutral:
it is invariant under $SO(6)$, $U(1)_R$, and the anomalous $U(1)$,
and is also dimensionless.
To calculate with this effective lagrangian, we must choose
independent fields to parameterize $M$ so that the constraint
\Eq{soconstr} is satisfied.
Expanding in these fields gives the terms in the effective lagrangian
in terms of derivatives of the function $k$.
The function $k$ is completely unknown, except that it must give
positive definite kinetic terms when expanded about any point.

Up to $SO(6)$ rotations the most general VEV can be written as
\beq\eql{MVEV}
\avg{M} = \pmatrix{(1 + v^2)^{1/2} \cr i v \cr 0 \cr \vdots \cr 0 \cr},
\eeq
where $-\infty < v < +\infty$ parameterizes the set of inequivalent
vacua in the SUSY limit.
For $v \ne 0$, $SO(6)$ is broken to $SO(4)$, while at $v = 0$ the
unbroken symmetry is enhanced to $SO(5)$.
We then write
\beq
M = \avg{M}(1 + \De) + \Phi,
\qquad
\avg{M_a} \Phi_a =0.
\eeq
The constraint \Eq{soconstr} can be solved to give
\beq
\De = -\sfrac{1}{2} \Phi^2 + O(\Phi^4).
\eeq
Using the basis
\beq
\Phi = \frac{1}{(1 + 2 v^2)^{1/2}}
\pmatrix{v \cr i(1 + v^2)^{1/2} \cr 0 \cr 0 \cr 0 \cr 0 \cr} \Phi_1
+ \pmatrix{0 \cr 0 \cr 1 \cr 0 \cr 0 \cr 0 \cr} \Phi_2
+ \cdots
+ \pmatrix{0 \cr 0 \cr 0 \cr 0 \cr 0 \cr 1 \cr} \Phi_5
\eeq
we have
\beq
\!\!\!\!\!\!\!\!
M^\dagger M = \avg{M^\dagger M}
+ \frac{2 v (1 + v^2)^{1/2}}{(1 + 2 v^2)^{1/2}} (\Phi_1 + \Phi_1^\dagger)
- \sfrac{1}{2} (\Phi - \Phi^\dagger)^2
+ O(\Phi^3).
\eeq
Note that $\avg{M^\dagger M} = 1 + 2 v^2$.

The vacuum energy as a function of $v$ can be determined from the
terms in the effective potential that are independent of the
scalar components of $\Phi$.
Eliminating the auxiliary components of $\Phi$, we obtain
\beq\eql{potdef}
V(v) = -[ \La^2_R ]_{\th^2\bar{\th}^2} \avg{k}
+{|[ \La^2_R ]_{\th^2}|^2\over [ \La^2_R ]_0}
{4v^2(1+v^2) \avg{k'}^2 \over (1+2v^2) \avg{k'} + 4v^2(1+v^2) \avg{k''}},
\eeq
where $\avg{k} = k(1 + 2 v^2)$, {\it etc\/}.
We do not know the function $k$ explicitly, but we know that
$k'$ must be nonzero everywhere on the moduli space in order for the
kinetic terms to be positive in the SUSY limit.
This is sufficient to conclude that the enhanced symmetry point $v=0$
is a local minimum for any positive soft scalar mass
($[ \La^2_R ]_{\th^2\bar{\th}^2} < 0$) and for any
gaugino mass ($[ \La_R ]^2_{\th^2}$).

In the absence of a gaugino mass only the first term in \Eq{potdef}
survives, so that the positivity of $k'$ is sufficient to conclude
that the global minimum is at $v = 0$.
In this case, the effective lagrangian is
\beq
\scr{L}_{\rm eff} = \myint d^2\th d^2\bar{\th}\,
\La_R^2 \, k'(1) \left[
\Phi^\dagger_a \Phi_a
- \sfrac{1}{2}( \Phi_a \Phi_a + \hc ) \right]
+ O(\Phi^4).
\eeq
Note that at this order the only dependence on the effective
\Kahler potential is through an overall factor, which cancels
when we compute masses.
The masses are
\beq
m^2_{\Re(\Phi)} = 0,
\qquad
m^2_{\Im(\Phi)} = +2 m_0^2,
\eeq
where $m_0^2$ is a common bare soft mass term of the fundamental
$Q$ fields.
The fields $\Re(\Phi)$
are massless because they are the Nambu--Goldstone
bosons of the global symmetry breaking $SO(6) \to SO(5)$.
(Local $SO(6)$ excitations of the VEV \Eq{MVEV} correspond to the real
components of $\Phi$.)
The mass of the $\Im(\Phi)$ fields is a simple multiple of the bare
mass, and is positive if the bare mass is positive.

These techniques can be applied to other models with deformed moduli
spaces, but we cannot generally determine the scalar masses.
Consider for example the case of $SU(N)$ gauge theory ($N \ge 3$)
with $N$ `flavors' of quarks $Q^j, \bar{Q}_{\bar{k}}$,
$j, \bar{k} = 1, \ldots, N$.
In the SUSY limit, the moduli space is parameterized by the gauge
invariants
\beq
M^j{}_{\bar{k}} \equiv \frac{1}{\La_S^2} Q^j \bar{Q}_{\bar{k}},
\quad
B \equiv \frac{1}{\La_S^N} \det Q,
\quad
\bar{B} \equiv \frac{1}{\La_S^N} \det \bar{Q} .
\eeq
The quantum constraint is
\beq\eql{qconstr}
\det M - B \bar{B} = 1.
\eeq
The most general effective lagrangian invariant under the
$U(N) \times U(N)$ flavor symmetry (which includes the anomalous
$U(1)$) and the anomaly-free $U(1)_R$ symmetry is
\beq
\scr{L}_{\rm eff} &= \myint d^2\th d^2\bar{\th}\, \La_R^2
\,k(B^\dagger, B, \bar{B}^\dagger, \bar{B}, M, M^\dagger)
+ \hbox{derivative terms},
\eeq
where $M$, $B$, and $\bar{B}$ satisfy the quantum constraint
\Eq{qconstr}.
(The content of the above equation is that $\La_R$ appears only as a
multiplicative factor.)

Suppose we are interested in the maximally symmetric point
$\avg{B} = \avg{\bar{B}} = 0$, $\avg{M} = 1$.
We write
\beq
M = \avg{M} + \Phi,
\eeq
and the constraint tells us that
\beq\eql{solvconstr}
\tr( \Phi) = B \bar{B} - \sfrac{1}{2} \tr(\Phi'^2) +
\hbox{\rm quadratic\ terms},
\eeq
where $\Phi' = \Phi - (\tr \Phi) / N$ is the trace-free part of $\Phi$.
No symmetry can forbid a term in the effective \Kahler potential of
the form
\beq\eql{badK}
k \sim B \bar{B} + \hc
\eeq
(A quick way to see this is that the combination $B \bar{B}$ appears
in the quantum constraint.)
The quantum constraint is solved by \Eq{solvconstr}, and so this term
is quadratic in terms of the independent fields.
Therefore, in the presence of soft masses, a term of the form \Eq{badK}
this term gives a `$B$ type' mass for the baryon fields, and we cannot
determine the masses for these fields.

Although it is of limited interest,
we can determine the mass-squared for the mesons in the maximally
symmetric vacuum.
The unbroken $U(N)$ `diagonal' symmetry means that the \Kahler potential
has the form
\beq
k = k_0 \left\{ \tr(\Phi'^\dagger \Phi') + c \left[
\tr(\Phi'^2) + \hc \right] + \cdots \right\}.
\eeq
The fact that there must be $N^2 - 1$ Nambu--Goldstone bosons from
the spontaneous symmetry breaking of the global symmetry means that
$c = -\sfrac{1}{2}$, and the soft masses for the meson fields are
determined.
As in the $SU(2)$ case, we find that the mass-squared for the
$N^2 - 1$ massive mesons is $+\sfrac{1}{2} m_{\phi 0}^2$.

\section{Anomaly-free Gauged $U(1)_R$}
In this Section we give a new derivation of the results of
\Ref{AR} reviewed in the previous Section that makes use of
\emph{anomaly-free} gauge symmetries.
This gives additional insight into why we are able to obtain exact
results for non-holomorphic quantities.
We use these results to obtain a very simple derivation of the
behavior of soft masses in a theory with a conformal fixed point.

The ideas are easiest to explain in the context of a $\scr{N} = 1$
supergravity (SUGRA) background with a gauged $U(1)_R$.
This can be formulated simply using the superconformal approach to
SUGRA \cite{sugra}.
In the flat limit, the tree-level lagrangian for a gauge theory
coupled to a SUGRA background can be written in superspace as
\cite{fgkv}
\beq\bal
\scr{L} &= \myint d^4\th\, (\phi^\dagger e^{-\frac{2}{3} V_R} \phi)
(Q^\dagger \scr{Z} e^V e^{V_R \cdot R} Q)
+ \left( \myint d^2\th\, \phi^3 W(Q) + \hc
\right)
\\
&\quad + \myint d^2\th\, S \tr(W^\al W_\al) + \hc
\eal\eeq
The chiral field $\phi$ is the conformal compensator,
whose role in the full formalism is to break the superconformal
symmetry down to super-Poincar\'e symmetry.
The $\phi$ dependence is completely determined by dilatation
invariance and $U(1)_R$ invariance, under which $\phi$ has respectively
weight $+1$ and charge $2/3$.
All other fields have vanishing weight and $U(1)_R$ charge.%
\footnote{In fact, it is worth noting that we do not need SUGRA
to determine the dependence of $\phi$.
However, SUGRA makes the dependence of $V_R$ more clear.}
In this approach to SUGRA, the $U(1)_R$ symmetry is part of the 
superconformal group, and is gauged (the gauge field is the SUGRA
vector auxiliary field).
The field $V_R$ is a gauge superfield for an \emph{ordinary} $U(1)$
gauge group whose charge matrix we call $R$.
The superconformal compensator is charged under this $U(1)$ with
$R_\phi = -\frac{2}{3}$.
The VEV of the conformal compensator $\phi = 1 + \cdots$ therefore
breaks $U(1) \times U(1)_R$ down to the diagonal $U(1)$ subgroup.
This unbroken group is an $R$ symmetry, and the matter fields have
charge $R$.
(This is the justification for the somewhat abusive notation used above,
where the charge of the ordinary $U(1)$ is denoted by $R$.)
It is this $U(1)$ symmetry that must be anomaly-free in order for the
dependence on $V_R$ to be fixed simply by considerations of charge
conservation.
As shown in \Ref{PR}, the condition that a $U(1)_R$ symmetry with
charges $R$ must be anomaly-free in order to define a consistent
deformation of the theory can also be derived without referring to
SUGRA.

We now consider the SUSY-breaking background
\beq
V_R = \th^2 \bar{\th}^2 D_R,
\qquad
\phi = 1 + \th^2 F_\phi.
\eeq
The field $D_R$ gives rise to soft masses at tree level, but the
dependence on $F_\phi$ is more subtle.
Note that if the lagrangian contains no dimensionful terms, then
$W(Q) \sim Q^3$ and the $\phi$ dependence can be completely
eliminated from the tree-level lagrangian by a field redefinition
$Q' = \phi Q$.
However, regulating the theory necessarily introduces mass parameters
and therefore brings in additional $\phi$ dependence at loop level
\cite{RS,GLMR}.
The coupling of $\phi$ is completely determined by dilatation symmetry,
so the loop effects are correctly included by the replacement
\beq
\mu^2 \to \hat{\mu}^2
= \frac{\mu^2}{\phi^\dagger e^{-\frac{2}{3} V_R} \phi}
\eeq
in the renormalized couplings $R(\mu) = 1/g^2(\mu)$ and
$\scr{Z}_r(\mu)$.
This gives rise to running scalar and gaugino masses \cite{PR}
\beq\eql{runscalar}
m^2_r(\mu) &= -[ \ln{\scr{Z}}_r(\hat{\mu}) ]_{\th^2 \bar{\th}^2}
= \left( \sfrac{2}{3} - R_r - \sfrac{1}{3} \ga_r \right) D_R
- \frac{1}{4} \frac{d \ga_r}{d\ln\mu} |F_\phi|^2
\\
\eql{rungaugino}
m_\la(\mu) &= [\ln R(\hat{\mu})]_{\th^2}
= \frac{\be(g^2)}{2 g^2} F_\phi,
\label{softr}
\eeq
where $\gamma_r=d\ln Z_r/d\ln\mu$ and $\be(g^2) = d g^2 / d\ln\mu$.
These equations define a consistent RG trajectory to all orders in
perturbation theory in an appropriate class of renormalization
schemes \cite{AGLR}.
The `bare' soft mass parameters on this RG trajectory are
\beq
m^2_{r 0} &= \lim_{\mu\to\infty} m^2_r(\mu)
= \left( \sfrac{2}{3} - R_r \right) D_R,
\\
\frac{m_{\la 0}}{g_0^2} &= \lim_{\mu\to\infty}
\frac{m_\la(\mu)}{g^2(\mu)}
= -\frac{b}{16\pi^2} F_\phi,
\eeq
where $b = 3 t_G - \sum_r t_r$ is the coefficient of the
gauge beta function.

For $SU(N_c)$ SUSY QCD with $N_f$ flavors,
\beq\eql{UVmass1}
m^2_{r 0} &= \frac{3 N_c - N_f}{3 N_f} D_R,
\\
\eql{UVmass2}
\frac{m_{\la 0}}{g_0^2} &= -\frac{3 N_c - N_f}{16\pi^2} F_\phi,
\eeq
We can now apply these results to the low-energy effective
theory to find the mapping of the UV soft masses onto
IR soft masses.
For $N_c + 1 < N_f < \sfrac{3}{2} N_c$ the low-energy
description has an infrared-free `dual' description in terms
of an $SU(N_f - N_c)$ gauge theory with dual quarks $q, \bar{q}$
and neutral `meson' fields $M$.
Because this theory is infrared-free, we can easily read off
the soft masses of these fields on the RG trajectory defined above
in the far infrared:
\beq
m^2_{q, {\rm IR}} &= \lim_{\mu\to 0} m^2_q(\mu)
= \left( \sfrac{2}{3} - R_q \right) D_R
= \frac{2 N_f - 3 N_c}{3 N_f} D_R,
\\
m^2_{M, {\rm IR}} &= \lim_{\mu\to 0} m^2_M(\mu)
= 2 \frac{3 N_c - 2 N_f}{3 N_f} D_R,
\\
\left. \frac{m_{\la_D}}{g_D^2} \right|_{\rm IR}
&= \lim_{\mu\to 0}  {m_{\la_D}(\mu) \over g_D^2(\mu)}
=-\frac{2N_f-3N_c}{16 \pi^2} F_\phi,
\eeq
where $\la_D$ is the dual gaugino and $g_D$ the dual gauge coupling.
Comparing to \Eqs{UVmass1} and \eq{UVmass2}, we obtain the
relation between the UV and IR soft masses obtained in
\Ref{AR}.
(The equations above are valid also in the s-confining case
$N_f = N_c + 1$, where the dual quarks are identified with the
baryons.)
Of course, the physical masses should be evaluated at a
renormalization scale $\mu$ equal to the physical mass.
However, this will give corrections to the masses of order
$g_D^2(\mu) / 16\pi^2$, where $g_D(\mu)$ is the running coupling
in the dual description.
These corrections are small if the dual description is weak.

The gauged non-anomalous $U(1)_R$ is also interesting for
theories with Yukawa couplings or multiple gauge factors. Here it
can be used to define some non-trivial, but nonetheless `integrable',
soft term RG flow.  Indeed the duals of pure gauge theories often
involve Yukawa couplings. The underlying $U(1)_R$ symmetry then makes it
more clear why in \Ref{AR} the soft term flow of the dual theory
could also be followed exactly.

We now consider SUSY QCD in the conformal window
$\sfrac{3}{2} N_c \le N_f \le 3 N_c$.
This was considered in \Ref{KKKZ}, where the explicit RG equations
of \Ref{jones} were used to show that all soft masses scale to zero.
In this approach, the origin of this result is clouded
in the computations; we believe that the supergravity approach gives
a significant clarification.

First, it is obvious that when the theory
approaches a scale invariant point the dependence on the scale
compensator $\phi$ must drop out from the effective action.
This is manifest in eq. \ref{softr}, since the contribution of $F_\phi$
is proportional  to $\be(g^2)$, which vanishes at the fixed point.
Second, if we choose $R$ to be the same for all quark fields,
the contribution of $D_R$ is proportional to
\beq
2 - 3 R_Q - \ga_Q \propto 3N_c - N_f - N_f \ga_Q,
\eeq
which is the quantity that controls the vanishing of the NSVZ
beta function:
\beq
\mu\frac{d g^2}{d\mu} = -\frac{g^4}{8\pi^2}\,
\frac{\displaystyle b - \sum_r t_r \ga_r}
{\displaystyle 1 - \frac{g^2}{8\pi^2} t_G}.
\eeq
This result holds because the $R$ charge chosen is identical with
the $U(1)_R$ charge in the superconformal algebra at the fixed point,
which satisfies $d_\scr{O} = \sfrac{3}{2} R_{\scr{O}}$ for any
chiral operator $\scr{O}$ with scaling dimension $d_{\scr{O}}$.
For example, the $\scr{O}$ 2-point function can be described by a term
in the 1PI effective action
\beq
\Ga_{\rm 1PI} \propto \myint d^4\th\,
\left( \O^\dagger e^{V_R R_{\scr{O}}} \Dal^{1-d_{\O}} \O \right)
\left (\phi^\dagger e^{-\frac{2}{3} V_R} \phi \right)^{d_\O}
+ \cdots
\eeq
in which the dependence on $V_R$ drops out.


These results give the scaling of the soft masses for
$\mu$ larger than the soft masses themselves;
below this scale, the soft masses are relevant perturbations
and the physics is no longer controlled by the fixed point.
The approach to the fixed point $g = g_*$ is given by
\beq
g^2(\mu) = g_*^2 + c \left( \frac{\mu}{\La} \right)^{\ga'},
\eeq
where $\La$ is the scale of strong interactions.
The critical exponent is
\beq
\ga' = \left( 1 - \frac{g_*^2}{8\pi^2} t_G \right)^{-1}
\sum_r \frac{g_*^2}{8\pi^2} t_r
\left. \frac{d\ga_r}{d\ln g^2} \right|_* > 0.
\eeq
In a strongly-coupled theory, \naive dimensional
analysis tells us that $\ga' \sim 1$.
By \Eqs{runscalar} and \eq{rungaugino} we find that the scaling of
soft terms is $m_Q(\mu)\sim F_\Phi (\mu/\Lambda)^{\ga'/2}$
and $m_\la(\mu) \sim F_\Phi (\mu/\Lambda)^{\ga'}$.
We see that, for $\mu \ll \La$,  $m_{Q} \gg m_{\la}$,
so the scalar masses control the exit from the fixed point.
Solving $m^2_Q(\mu)(\mu \sim m_Q) \sim m_Q$ gives
\beq
\frac{m_Q}{\La} \sim
\left( \frac{F_\phi}{\La} \right)^{2 / (2 - \ga')},
\qquad
\frac{m_\la}{\La} \sim
\left( \frac{F_\phi}{\La} \right)^{(2 + \ga')/(2 - \ga')},
\eeq
where we have assumed that the gaugino masses essentially freeze upon
exiting the fixed point.
For $\ga' > 2$, this solution is not applicable.
In that case, the scalar mass is scaling to zero faster
than $\mu$ itself, and the physical soft masses vanish.
This is a logical possibility in strongly-coupled theories, but
unfortunately we are unable to compute $\ga'$ and so we cannot
determine whether this occurs.%
\footnote{The anomalous dimension $\ga'$ can be calculated
in weakly coupled theories, such as the $1/N_c$ expansion
around the Banks-Zaks fixed point at
$3N_c-N_f=\epsilon N_c$ in SQCD.
There one finds $\ga'\sim \epsilon^2 \ll 1$ \cite{KKKZ}.
The question seems to be open whether $\ga' > 2$ in
the middle of the conformal window.}

We close this Section with some remarks on the possible
phenomenological applications of conformal theories.
The fact that soft masses decrease as a non-trivial power law
in the infrared in stronlgly-coupled conformal theories
raises the possibility that this could play a role in understanding
the smallness of SUSY breaking in our world.
However, there are some very generic difficulties with this idea.
First, as pointed out above, the gaugino masses are always smaller
than the scalar masses in such a scenario.
Second, the reduction of the scalar mass discussed above applies
only to the component proportional to the anomaly-free
$U(1)_R$ generator.
All flavor breaking scalar masses associated to the anomaly-free generators
of the flavor group ($SU(N_f) \times SU(N_f)$ for SQCD)
will not undergo the suppression discussed above.
Since realistic supersymmetric theories require the squark masses to
be approximatively flavor-preserving, this will make the SUSY flavor
problem more severe.
However, strongly-coupled theories near their conformal fixed
points may play a role in nature for other reasons, and it is
important to know how the soft masses scale in such
theories.

\section{${\cal N} = 2$ Super Yang--Mills}
We now turn our attention to $SU(2)$
$\scr{N} = 2$ super Yang--Mills theory.

\subsection{${\cal N} = 2$ Spurion Analysis}
We first consider the theory formulated in ${\cal N} = 2$
superspace and perform a spurion analysis by generalizing the couplings
to $\scr{N} = 2$ superfields.
This analysis generalizes the results of \Ref{ADKM} because we use more
general ${\cal N} = 2$ spurions.
This is sufficient to parameterize all soft SUSY breaking except for a
non-holomorphic soft mass for the scalar field.
We work out the effects of this breaking on the low-energy potential
using $\scr{N} = 2$ techniques and compare our results to those of
\Ref{ADKM}.
We also give an elementary derivation of the
relation between the modulus and the prepotential.

The gauge multiplet can be described in ${\cal N} = 2$ superspace
$(x^\mu, \th^\al, \tilde{\th}^\al, \th^\dagger_{\dot\al},
\tilde{\th}_{\dot\al}^\dagger)$
by a superfield ${\cal A}$ satisfying \cite{GSW}
\beq\eql{chiral}
\bar{D}^j_{\dot\al}{\cal A} &= 0,
\\
\eql{vector}
D^i D^j {\cal A} &= \bar{D}^i \bar{D}^j \bar{\cal A},
\eeq
where $i,j$ are $SU(2)_R$ indices
($\th^\al_1 = \th^\al$, $\th^\al_2 = \tilde{\th}^\al$).%
\footnote{We use the conventions of \Ref{GSW}, which extend the conventions
of Wess and Bagger \cite{WB} to $\scr{N} = 2$ superspace.
For a precise definition of $\bar{\scr{A}}$, see \Ref{GSW}.}
Notice that by $SU(2)_R$ covariance $\bar D^i=\epsilon^{ij}(D^j)^*$, where
$\epsilon^{ij}$ is the antisymmetric tensor.
\Eq{chiral} states that $\scr{A}$ is a $\scr{N} = 2$ chiral multiplet,
while \Eq{vector} is a reality condition that defines an
${\cal N} = 2$ vector multiplet.
In this notation, the lagrangian is
\beq
\scr{L} = \myint d^2\th d^2\tilde{\th}
\frac{1}{2 g^2}
\tr({\cal A}^2) + \hc
\eeq
(We have set the theta term to zero.
We take ${\cal A}={\cal A}_b \tau_b$
with the $SU(2)$ generators normalized by
$\tr (\tau_a\tau_b)= \frac{1}{2} \delta_{ab}$.)
The $\scr{N} = 1$ decompositions of $\scr{A}$ is
\beq
\scr{A} &= \Phi
+ i\sqrt{2} \tilde{\th}^\al W_\al
+ \tilde{\th}^2 \left[ -\sfrac{1}{4} \bar{D}^2
\left( e^V \Phi e^{-V} \right) \right],
\eeq
where $\Phi$ and $W_\al$
are $\scr{N} = 1$ chiral superfields
that are functions of $\th$ and
\beq
y^\mu = x^\mu + i(\th \si^\mu \th^\dagger
+ \tilde{\th} \si^\mu \tilde{\th}^\dagger).
\eeq

We now consider extending the gauge coupling to a $\scr{N} = 2$
superfield:
\beq\eql{NtooL}
\scr{L} = \myint d^2\th d^2\tilde{\th}\, \Si
\tr({\cal A}^2) + \hc
\eeq
This is $\scr{N} = 2$ supersymmetric provided that
$\Si$ is chiral:
\beq
\bar{D}^j_{\dot\al} \Si = 0.
\eeq
\Ref{ADKM} also performed a $\scr{N} = 2$ spurion analysis, but they
imposed the additional condition that $\Si$ is a $\scr{N} = 2$ vector
multiplet.

The low-energy effective theory arising from the strong dynamics
depends on $\Si$ through the RG-invariant scale
\beq
\La = \mu e^{-4\pi^2 \Si(\mu)},
\eeq
where we have used the beta function appropriate for $SU(2)$.
(Recall that $\scr{N} = 2$ SUSY implies that the gauge coupling runs only
at one loop.)
Note that this is an $\scr{N} = 2$ chiral superfield.
Away from the monopole points, the effective theory can be written in
terms of a $U(1)$ gauge superfield $a$:
\beq\eql{ntooleff}
\scr{L}_{\rm eff} = \frac{1}{4\pi} \Im \myint d^2\th d^2\tilde{\th}\,
\scr{F}(a, \La),
\eeq
where dimensional analysis implies
\beq\eql{littlef}
\scr{F}(a, \La) = \La^2 \scr{G}(a/\La).
\eeq

These considerations lead directly to an elementary proof of the relation
between the modulus $u \equiv \tr\phi^2$ and the prepotential.
Suppose we turn on a $\th^2 {\tilde\th}^2$ component of
$\Si$ as a source:
\beq
\Si = \frac{1}{2 g^2} + \th^2 \tilde{\th}^2 \tilde{D}.
\eeq
    From the fundamental lagrangian, we see that
\beq
\left. \frac{\de\Ga_{\rm 1PI}}{\de \tilde{D}} \right|_{\tilde{D} = 0}
= \avg{u}.
\eeq
In the effective theory, we can evaluate the term linear in $\tilde{D}$
in $\de\Ga_{\rm 1PI}/ \de\tilde{D}$ by expanding out the $\th$- and
$\tilde{\th}$-dependent terms in $\La$:
\beq
\tilde{D} \avg{u} &= \frac{1}{8\pi i} \left[ \La^2 \scr{G}(a/\La)
\right]_{\th^2\tilde{\th}^2}
\nonumber\\
&= \frac{1}{8\pi i} \left[
[\La^2]_{\th^2 \tilde{\th}^2} \scr{G}(a / \La)
+ \La^2 \scr{G}'(a/\La) \cdot a [1/\La]_{\th^2 \tilde{\th^2}} \right]
\nonumber\\
&= \frac{\pi}{2i} \left[ \tilde{D} \left( 2 \scr{F} - a a_D \right) \right],
\eeq
where $a_D \equiv \partial\scr{F} / \partial a$.
This immediately gives
\beq\eql{matone}
\scr{F} - \frac{1}{2} a a_D = -\frac{i}{\pi} u.
\eeq
Previous derivations \cite{matone} of this result have relied on specific
properties of the Seiberg--Witten solution.
Here we see that it follows from elementary spurion considerations.

We can also use this formalism to work out the effects of soft SUSY
breaking in the low-energy theory.
If we write
\beq
\Si = \frac{1}{g^2} \left[ \sfrac{1}{2} + \th^2 m_\la
+ \tilde{\th}^2 m_\chi + 2 i \th\tilde{\th} m_D
- \th^2 \tilde{\th}^2 m_B^2 \right]
\eeq
then this gives rise to SUSY breaking terms in the fundamental
lagrangian:
\beq\eql{softuv}
\De\scr{L} =& \frac{1}{g^2} \tr\left[
-m_\la \la \la - m_\chi \chi \chi
- 2i m_D \la\chi - (2g^4\Delta +m_B^2 )\phi^2
\right] + \hc\nonumber\\
&-2g^2T\tr(\phi\phi^*)
\eeq
where
\beq
T = \frac{1}{g^4} \tr(m_\Psi^\dagger m_{\Psi}),
\qquad
\De = \frac{1}{g^4} \det(m_\Psi),
\qquad
m_{\Psi} = \pmatrix{m_\chi & -im_D \cr -im_D & m_\la \cr}.
\eeq
and where $\la$ and $\chi$ are the fermion components of $\scr{A}$.
(Because the kinetic terms are multiplied by a factor of $1/g^2$,
the mass parameters as defined here are the running masses. In ${\cal N}=2$
there is no running beyond 1-loop, or, equivalently, the holomorphic
and 1PI coupling can be take to coincide, $R=S+S^\dagger$.
As a reflection of that the matrix $m_{\Psi}/g^2$ is RG invariant,
and coincides with
the bare parameter at $\mu=\infty$ defined in Section 2.)
We can now work out the effects of this soft SUSY breaking in the
low-energy theory directly from the Seiberg-Witten solution for
$\scr{G}$ in \eq{littlef} by expanding out the $\th$- and
$\tilde{\th}$-dependent terms in $\La$.

Specifically, the terms relevant for the potential for the effective
theory are
\beq
a = \si + \th^2 f + \tilde{\th}^2 f^\dagger + i \th \tilde{\th} d,
\eeq
where $\si$ is the complex propagating scalar, and $f$, $d$
are auxiliary fields.
The reality condition on $a$ implies that $d$ is real, relates the
coefficients of the $\th^2$ and $\tilde{\th}^2$ terms, and implies the
absence of a $\theta^2\tilde \theta^2$ component.
The effective lagrangian including soft SUSY breaking is then simply
given by \Eq{ntooleff}, where we expand the $\th$- and $\tilde{\th}$
dependence of both $a$ and $\La$.
After some straighforward algebra (and use of \Eq{matone}), we obtain
\beq
\eql{pot}
V_{\rm eff} = \frac{T}{k_{u^\dagger u}}
+8\pi^2 \Re \left[ \De \cdot \left(
2 u - \frac{k_{u^\dagger}}{k_{u^\dagger u}} \right) \right]
+ \frac{2}{g^2} \Re (m_B^2 u),
\eeq
Note that $SU(2)_R$ is manifest.
The results above are valid away from the monopole points;
we will postpone the discussion of the physics of this result to the
next Section, where we consider the most general soft breaking terms.

Before we leave the subject of $\scr{N} = 2$ spurions, we comment on the
relation between our results and those of \Ref{ADKM}.
In that paper the `dilaton' $S = i\Sigma$ is taken to be a vector superfield.
This corresponds to setting $m_B^2 = 0$, and $m_\lambda = -m_\psi^\dagger$
with $m_D$ pure imaginary, which gives $T = -2\Delta$.
$SU(2)_R$ invariance means that the perturbation is to just one independent
soft parameter $T$.
    From \Eq{softuv} it is manifest that for this choice of parameters
the half-line $\tr\phi^2 = \tr(\phi\phi^\dagger) > 0$
remains flat to all orders in perturbation theory.
Non-perturbative effects remove this flatness and the vacuum
is picked out along this half-line at the monopole point $u=1$ \cite{ADKM}.
Notice that if we had taken
$\Sigma$ rather that $i\Sigma$ to be vectorlike, the flat direction would
have been along $\tr\phi^2 = -\tr(\phi\phi^\dagger) < 0$, and the vacuum
would be stabilized at the dyon point $u = -1$.
This explains the apparent asymmetry between the monopole and dyon
points in \Ref{ADKM}.%
\footnote{%
In fact, under the $Z_8$ $R$-symmetry of the theory, we have $u \mapsto -u$,
$\Sigma(\th, \tilde\th) \mapsto \Sigma(\th e^{i\pi/4}, \tilde\th e^{i\pi/4})$.
Therefore the symmetry that exchanges monopole
and dyon poins is $m_\Psi \to i m_\Psi$,
$(T,\Delta) \mapsto (T,-\Delta)$.}

\subsection{General Spurion Analysis}
We now consider the most general soft breaking down to $\scr{N} = 0$.
In $\scr{N} = 1$ superspace, the action including the most general soft
breaking terms can be written
\beq\bal
\scr{L} &= \myint d^2\th d^2\bar{\th}\,
\scr{Z}_\Phi \tr( \Phi^\dagger e^V \Phi e^{-V} )
+ \left( \myint d^2\th\, S \tr( W^\al W_\al ) + \hc \right)
\\
&\qquad
+ \left( \myint d^2\th\, m \tr(\Phi^2) + \hc \right),
\eal\eeq
where $\scr{Z}_\Phi$, $S$, and $m$ are now regarded as
superfields with $\th$-dependent
components parameterizing the SUSY breaking.
Because $\Phi$ transforms as an adjoint under the gauge group,
there is an additional allowed soft term of the form
\beq
\eql{adddsoft}
\myint d^2\th\, \th^\al \tr(W_\al \Phi) + \hc,
\eeq
which gives rise to a mixing mass between the gaugino and $\Phi$ fermion.%
\footnote{This term also gives a term in the scalar potential proportional
to $\tr( \phi [\phi^\dagger, \phi]) + \hc$ that vanishes identically.}
However, the theory has a $SU(2)_R$ symmetry that is not manifest in the
$\scr{N} = 1$ formulation under which the gaugino and $\Phi$ fermion
form a doublet.
We therefore choose the `$\scr{N} = 1$'
direction to diagonalize the fermion mass matrix
and eliminate the term \Eq{adddsoft}.
Under $SU(2)_R$, the fermions $\chi$ and $\la$ form a doublet
\beq
\Psi^j = \pmatrix{\chi \cr \la \cr},
\eeq
and the fermion masses form a triplet
\beq
(m_{\Psi 0})_{jk} = \pmatrix{m_{\chi 0} & 0 \cr 0 & m_{\la 0} \cr}.
\eeq
The $SU(2)_R$ invariance of our results will be a non-trival check
of our formalism.

In the $\scr{N} = 2$ SUSY limit, we have $\scr{Z}_\Phi \to S + S^\dagger$,
and therefore $R \to S + S^\dagger$, $\La_R^2 \to |\La_S|^2$.
We now add the most general soft SUSY breaking terms:
a scalar mass $m^2_\phi$ for the scalar component of $\Phi$;
a $B$-type mass term $B \tr\phi^2 + \hc$;
and fermion masses $m_\la$ and $m_\chi$, where $\la$ is the gaugino
and $\chi$ is the fermion component of $\Phi$.
As discussed above, these terms can be viewed as $\th$-dependent
terms in the superfield coupling constants.
As might be expected, this technique is especially powerful in
theories with $\scr{N} = 2$ SUSY.
For example, using the gauged $U(1)_R$ or the results of Section 2,
one finds
\beq
\scr{Z}_\Phi(\mu) = \hat\scr{Z}_\Phi R(\mu)
= \hat\scr{Z}_\Phi \left (S(\mu)+S^\dagger(\mu) -
{1\over 4\pi^2}\ln\hat\scr{Z}_\Phi \right )\eeq
where $\hat\scr{Z}_\Phi$ is the RG invariant wave function of \Eq{wave}.
In the $U(1)_R$ approach one has $e^{-2V_R/3}= \hat\scr{Z}_\Phi$.
This gives a simple exact closed-form expression for the running soft
parameters.

The bare parameters $m_{\phi0}^2$ and $m_{\lambda 0}$ are defined
as in Section 2.
Moreover we define in the same spirit
\beq
m = \frac{m_{\chi 0}}{g_0^2} - \th^2 \frac{m_{B 0}^2}{g_0^2}
\eeq
in a field basis where $[\ln\scr{Z}_\Phi]_{\th^2} = 0$.
%

As long as the soft SUSY breaking parameters are small in units of
$\La$, they can be treated as a perturbation on the strong dynamics.
These lift the flat directions and give a potential on the
moduli space of SUSY vacua that we will determine.
The moduli space can be parameterized by the chiral gauge-invariant
operator
\beq
u \equiv \frac{1}{\La_S^2} \tr(\Phi^2).
\eeq
Note that $\La_S$ and $\Phi$ transform in the same way under the
anomalous $U(1)$, so $u$
is completely neutral: it is dimensionless, and uncharged under
the anomaly-free $U(1)_R$ symmetry
\beq
\Phi(\th) \mapsto \Phi(\th e^{-i\al}),
\quad
V(\th) \mapsto V(\th e^{-i\al}),
\quad
m(\th) \mapsto e^{2i\al} m(\th e^{-i\al}),
\eeq
as well as the anomalous $U(1)$ symmetry
\beq\eql{anomu1}
\Phi \mapsto e^A \Phi,
\quad
V \mapsto V,
\quad
\La_S \mapsto e^{A} \La_S,
\quad
m \mapsto e^{-2A} m.
\eeq
Neutral variables similar to $u$
are also present in the $\scr{N} = 1$ theories with
deformed moduli spaces discussed in Section 2.2.
This is no accident, since the moduli space is in a sense `deformed'
in the Seiberg--Witten solution, allowing the holomorphic prepotential
to be a nontrivial meromorphic function of $u$.

\subsection{Away from the Monopole Points}
We begin by describing the theory away from the monopole/dyon
points $\avg{u} = \pm 1$.
As long as $|\avg{u} - (\pm 1)| \gsim 1$, the only light (compared to
$\La$) states in the theory are the $U(1)$ gauge multiplet and the
modulus field $u$.
The most general effective lagrangian compatible with $\scr{N} = 1$
supersymmetry, holomorphy, and anomalous $U(1)$ invariance is
\beq\bal
\scr{L}_{\rm eff} &= \myint d^2\th d^2\bar{\th}\,
\La_R^2 k(u^\dagger, u)
+ \left( \myint d^2\th\, \sfrac{1}{2} s(u) w^\al w_\al
+ \hc \right)
\\
&\qquad + \left( \myint d^2\th\,
\La_S^2 m u + \hc \right)
+ \hbox{derivative terms},
\eal\eeq
where $w_\al$ is the $U(1)$ gauge field strength.
Note that all SUSY breaking is contained in the spurions $\La_R$
and $\La_S$, as follows purely from ${\cal N} = 1$ reasoning.

The \Kahler function $k$ cannot be determined from $\scr{N} = 1$
considerations, but it is completely fixed by the Seiberg--Witten solution
in the $\scr{N} = 2$ SUSY limit.
It is crucial for our results that the Seiberg--Witten solution also
determines the purely chiral (or antichiral) part of the \Kahler
potential.
(Recall that our inability to fix such terms was responsible for our
inability to determine the vacuum in $\scr{N} = 1$ theories with deformed
moduli space discussed previously.)
These terms vanish in the SUSY limit,
but they contribute to the potential when SUSY is broken explicitly.
In the SUSY limit, these terms can be probed by
promoting the $\scr{N} = 2$ gauge coupling superfield to a dilaton
source, as done in \Ref{ADKM}.
They are then fixed by the modular invariance of the Seiberg--Witten
solution in the presence of the dilaton, which ensures that if we travel
around a closed path in the moduli space we return to the same theory up to
a duality transformation.
Adding a chiral plus antichiral term to the
effective superpotential corresponds to modifying the $\scr{N} = 2$
prepotential by
$\scr{F}(a) \to \scr{F}(a) + \hbox{\rm const} \times a$,
but this clearly breaks modular invariance.
(More direct physical arguments that the linear terms in the
prepotential are fixed are also given in \Ref{SW}.)

Combining this with the expressions for the higher components of
$\La_R$ given in \Eqs{LaRthth} and \eq{LaRthththth}, we can
compute the potential for the scalar component of $u$.
The result is
\beq\bal
\!\!\!\!\!\!\!\!\!\!\!\!
V &= |\La|^2 \left(
m_{\phi 0}^2 - \left| \frac{4 \pi^2 m_{\la 0}}{g_0^2} \right|^2
\right) k
- \left[ \La^2 \left( \frac{m_{B0}^2}{g_0^2}
- 2 \frac{4\pi^2 m_{\chi 0} m_{\la 0}}{g_0^4} \right) u + \hc \right]
\\
&\quad
+ |\La|^2 \left| \frac{4\pi^2 m_{\la 0}}{g_0^2} \right|^2
\frac{| k_{u} |^2}{k_{u^\dagger u}}
-\left( \La^2 \frac{4\pi^2 m_{\chi 0} m_{\la 0}}{g_0^4}
\frac{k_{u^\dagger}}{k_{u^\dagger u}}
+ \hc \right)
+ \frac{|\La|^2 |m_{\chi 0}|^2}{k_{u^\dagger u}},
\eal\eeq
where $k_u = \partial k / \partial u$, \etc
Here
\beq
\La = \mu e^{-4\pi^2 / g^2(\mu)} e^{i\Th / 4}
\eeq
is the holomorphic scale in the $\scr{N} = 2$ limit.
One can check
that the dependence on the vacuum angle $\Th$ and the fermion masses
is the correct one dictated by the chiral anomaly.
The result above does not have manifest $SU(2)_R$
symmetry because the coefficients of $|m_{\la 0}|^2$ and $|m_{\chi 0}|^2$
are different.
(Note that $m_{\la 0} m_{\chi 0} = \det(m_{\Psi 0})$
is $SU(2)_R$ invariant.)
The potential is $SU(2)_R$ invariant if and only if
\beq\eql{Matprime}
k k_{u^\dagger u}
- \left| k_{u} \right|^2
= -\frac{1}{(4\pi^2)^2}.
\eeq
This relation is in fact satisfied, as can be seen by differentiating the
relation \Eq{matone}.
Since our formalism is not manifestly $SU(2)_R$ invariant, this
is a non-trivial check.
Using \Eq{Matprime} we can simplify the expression for the
potential to obtain
\beq\eql{theV}
V = \frac{\La^2}{4\pi^2} \sum_{j = 1}^6 m^2_{{\rm soft},j}
f_j(u^\dagger, u),
\eeq
where
\begin{eqaligntwo}
m^2_{{\rm soft},1} &= m_{\phi 0}^2,
&
f_1 &= 4 \pi^2 k,
\nonumber\\
m^2_{{\rm soft},2} &= 4\pi^2 T_0,
&
f_2 &= \frac{1}{4 \pi^2 k_{u^\dagger u}},
\nonumber\\
\eql{thef}
m^2_{{\rm soft},3} &= 4\pi^2 \Re \left(\De_0\right),
&
f_3 &= 8\pi^2 \Re
\left( 2u - \frac{k_{u^\dagger}}{k_{u^\dagger u}} \right),
\\
m^2_{{\rm soft},4} &= 4\pi^2 \Im \left(\De_0\right),
&
f_4 &= -8 \pi^2 \Im
\left( 2u - \frac{k_{u^\dagger}}{k_{u^\dagger u}} \right),
\nonumber\\
m^2_{{\rm soft},5} &=
4\pi^2 \Re \left(
\frac{m_{B 0}^2}{g_0^2} \right),
&
f_5 &= 2 \Re(u),
\nonumber\\
m^2_{{\rm soft},6} &=
4\pi^2 \Im \left(
\frac{m_{B 0}^2}{g_0^2} \right).
&
f_6 &= -2 \Im(u),
\nonumber
\end{eqaligntwo}
where
\beq
T_0 \equiv {1\over g_0^4}\tr |m_{\Psi 0}|^2,
\qquad
\De_0 \equiv {1\over g_0^4}\det(m_{\Psi 0}).
\eeq
These results agree with the results of the previous Subsection
obtained using $\scr{N} = 2$ arguments.%
\footnote{As already mentioned in Section 4.1, the soft terms induced by
an ${\cal N}=2$ dilaton spurion and studied in \Ref{ADKM} correspond
to the choice
$m_{\rm soft,2}^2 = -2 m_{\rm soft,3}^2$ with all other soft terms
vanishing.
The ${\cal N} = 1$ mass perturbation studied in \Ref{SW} simply
corresponds to $m_{\rm soft,2}^2 \ne 0$ with all other terms vanishing.}
We have factored out powers of $4\pi^2 = 16\pi^2 / b$ according to the
expectations of \naive dimensional analysis \cite{NDA}.
The overall factor of $1/(4\pi^2)$ arises because the potential is
a 1-loop effect, while the factors of $4\pi^2$ in the definitions of
$m_{{\rm soft},j}^2$ are chosen so that these quantities are equal to
soft parameters renormalized at the scale $\La$ where the theory
becomes strong: $g^2(\La) \sim 4 \pi^2$.
If \naive dimensional analysis is reliable, then the functions
$f_j$ should all be order 1.

The functions $f_1, \ldots, f_4$ are plotted in Figs.~1--4.
There are several interesting points to note about the results.
First, note that the functions $f_1, \ldots, f_4$ are all of order
1, as predicted by \naive dimensional analysis.
This is striking evidence for the correctness of these ideas
in the context of supersymmetric theories.

Although the results we have derived are not justified near the
monopole/dyon points, the behavior near these points is interesting.
Note that $m^2_{\phi 0}$ and $\Im(\De_0)$ apparently
drive the theory away from the monopole/dyon points,
while $T_0$ drives the theory toward the monopole/dyon points.
$\Re(\De_0)$ apparently gives a local minimum at either
the monopole or dyon points, depending on the sign.
When we will consider the theory near the monopole points, we
will find that these conclusions are in fact correct.

Finally, note that the results above predict a rich phase structure
as the various soft breaking terms are varied.
To give only one example, it can be seen that there is a first-order
phase transition between a Coulomb and a confined phase as we
increase the ratio $m^2_{\Phi 0} / T_0$.

\begin{figure}[p]
\centerline{\epsfxsize=5.0in\epsfbox{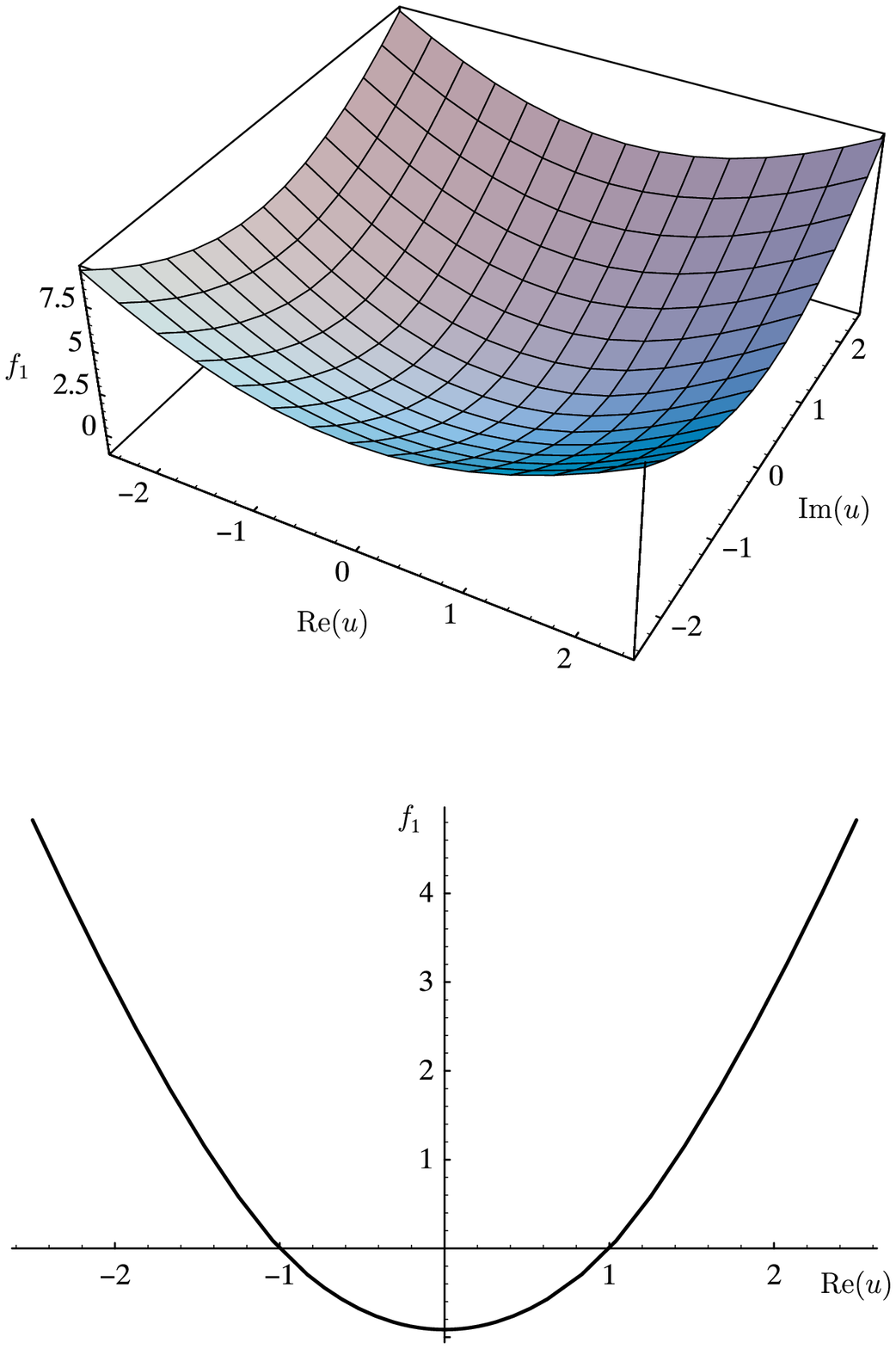}}
\smallskip
\caption{Fig.~1.\ \ Potential induced by a soft scalar mass
$m^2_{\phi 0}$.}
\end{figure}

\begin{figure}[p]
\centerline{\epsfxsize=5.0in\epsfbox{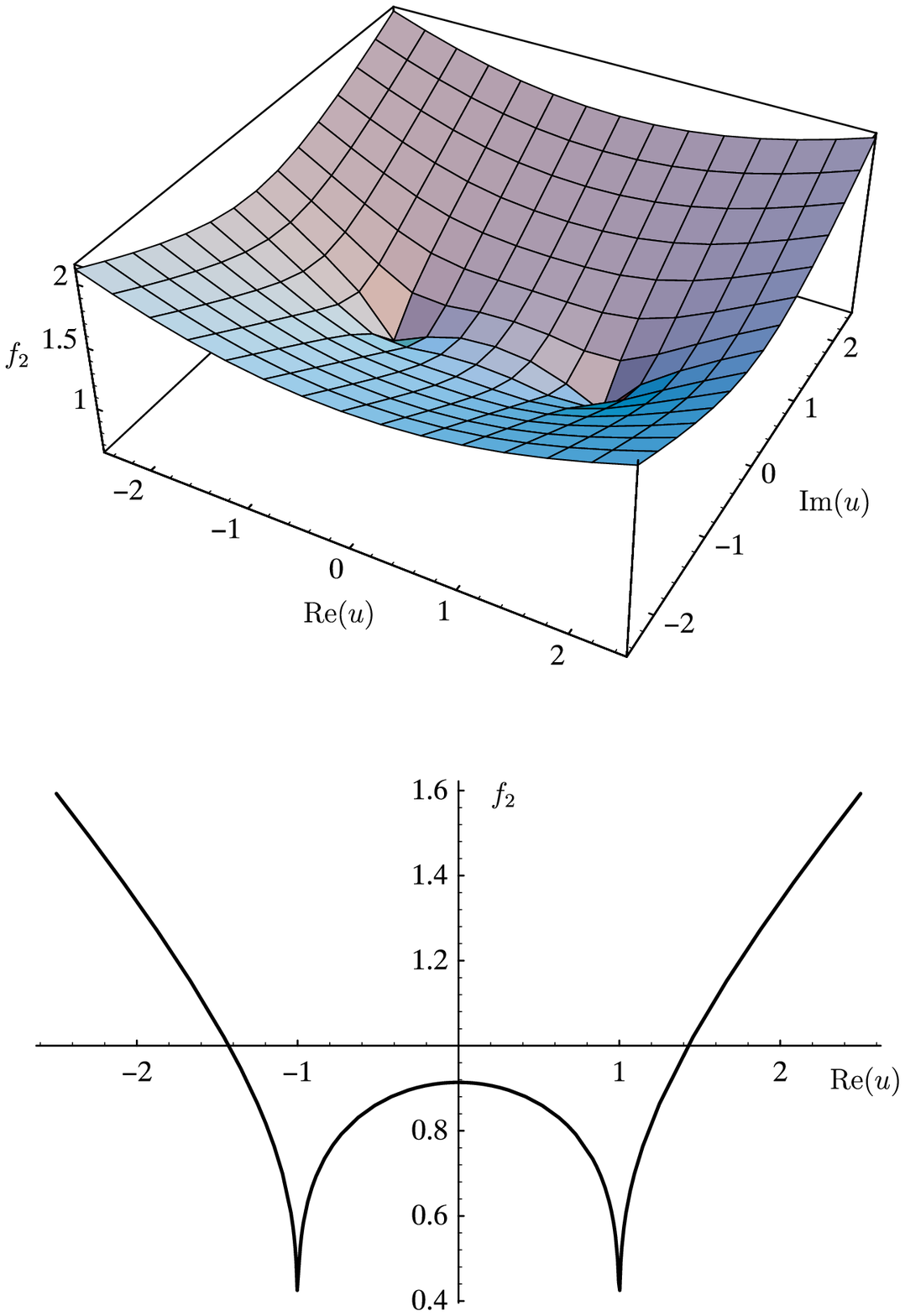}}
\smallskip
\caption{Fig.~2.\ \ Potential induced by the trace of the fermion
mass matrix $T_0$.
The potential approaches a finite value at the cusp singularities 
at the monopole/dyon points $u = \pm 1$.}
\end{figure}

\begin{figure}[p]
\centerline{\epsfxsize=5.0in\epsfbox{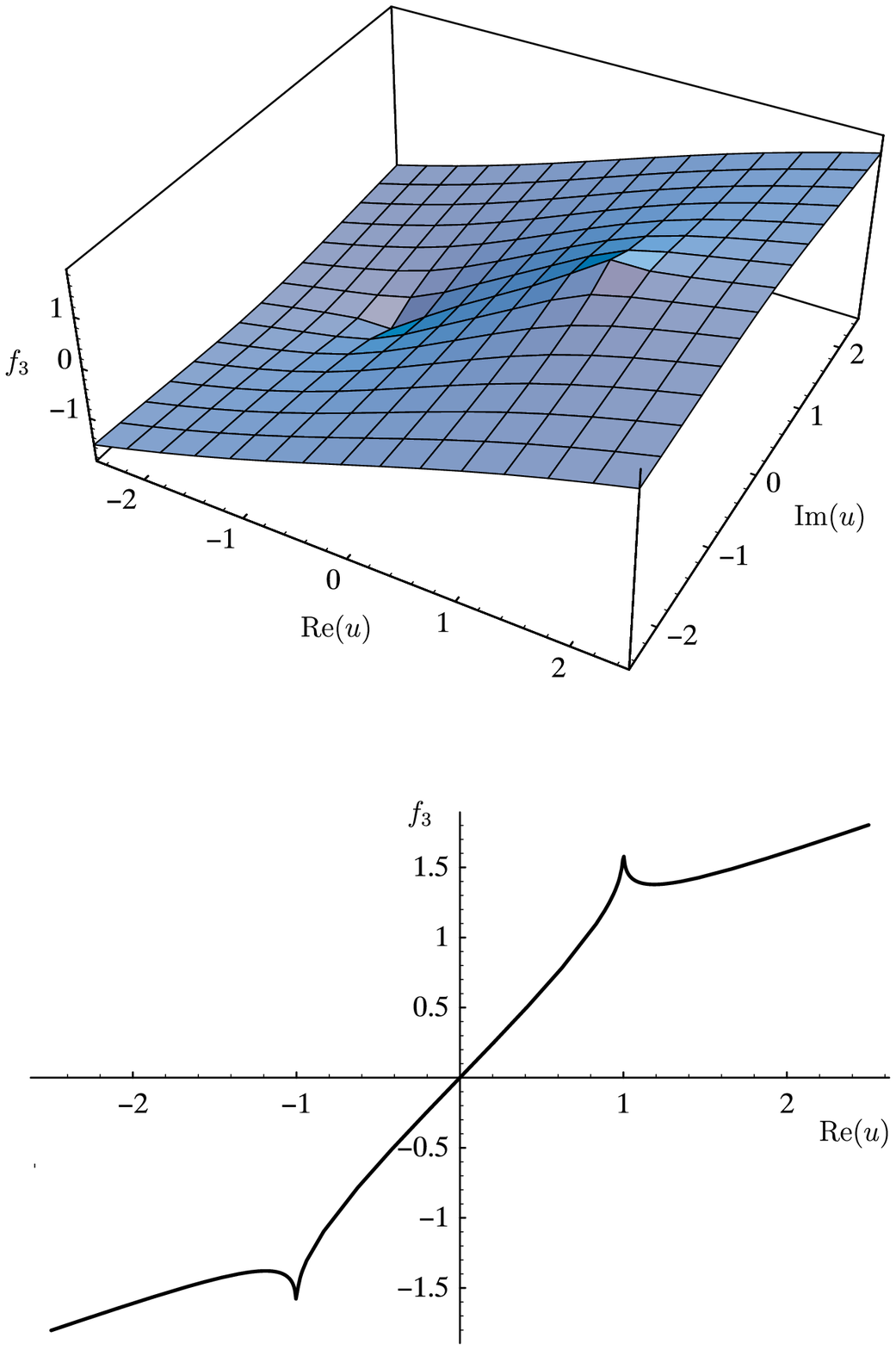}}
\smallskip
\caption{Fig.~3.\ \ Potential induced by $\Re(\De_0)$, where
$\De_0$ is the determinant of the fermion mass matrix.
The potential approaches a finite value at the cusp singularities 
at the monopole/dyon points $u = \pm 1$.}
\end{figure}

\begin{figure}[p]
\centerline{\epsfxsize=5.0in\epsfbox{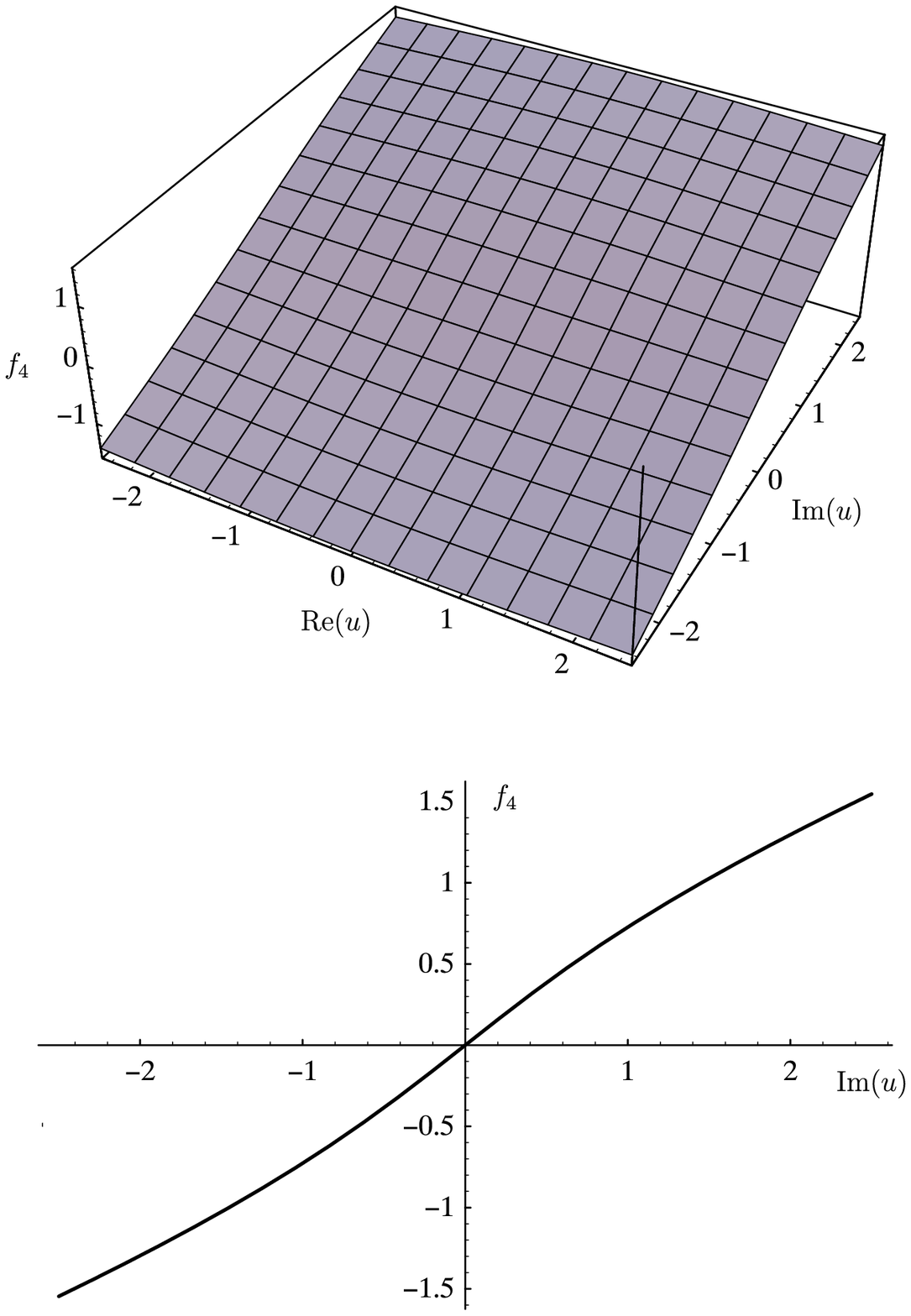}}
\smallskip
\caption{Fig.~4.\ \ Potential induced by $\Im(\De_0)$, where $\De_0$
is the determinant of the fermion mass matrix.}
\end{figure}

We now consider the question of how close we can get to the monopole
points $u = \pm 1$ before the results above break down.
The reason that the effective theory breaks down near the monopole points
is that there are extra monopoles (or dyons) with mass
\beq
m_M \sim \La [ \avg{u} - (\pm 1) ].
\eeq
The Seiberg--Witten solution away from the monopole points gives the
\emph{exact} effective lagrangian (up to higher derivative terms) with the
monopole integrated out.%
\footnote{Higher derivative terms affect the scalar potential at 
${\cal O}(m_{\rm soft}^3)$, while the leading effects we compute are order
$m_{\rm soft}^2$.
Therefore, higher derivative terms are negligible for small
$m_{\rm soft}$.}
Therefore, as long as $m_M \gg m_{\rm soft}$, it is a good
approximation to integrate out the monopoles.
This means that the results above are valid as long as
\beq
|\avg{u} - (\pm 1)| \gg \frac{m_{\rm soft}}{\La}.
\eeq
The corrections are suppressed by powers of $m_{\rm soft} / m_M$ for
$|\avg{u} - (\pm 1)| \lsim 1$, and are of order
$m_{\rm soft} / \La$ for $|\avg{u} - (\pm 1)| \gsim 1$.
For $m_{\rm soft} \ll \La$ this means that we can trust the above results
up to  a small region $|u - 1|\lsim m_{\rm soft}/\La$.
Inside this region the monopole VEV's can be turned on and decrease the
energy.
However we will later show that this effect is parametrically small
and it does not significantly alter the picture of where the vacuum
resides.
We can therefore conclude that
$m_{\phi 0}^2$ and $m_{B 0}^2$ push the vacuum away from the monopole
points, while fermion masses $m_{\la 0}$ and $m_{\chi 0}$ tend to
stabilize the monopole points.

\subsection{Near the Monopole Points}
We now describe the effective theory near the monopole point
$\avg{u} = +1$.%
\footnote{The point $\avg{u} = -1$ is related to this point
by charge conjugation.}
%
%
As argued in \Ref{SW},
near $\avg{u} = +1$ the weakly-coupled light degrees of freedom are
the modulus field $u$, the \emph{dual} photon field $v_D$,
and the monopole fields $M$, $\bar{M}$.
The effective lagrangian is therefore
\beq\eql{mleff}
\bal
\scr{L}_{\rm eff} &= \myint d^2\th d^2\bar{\th}\,
\La_R^2 k_D(u^\dagger, u)
+ \left( \myint d^2\th\, \sfrac{1}{2} s_D(u) w_D^\al w_{D \al}
+ \hc \right)
\\
&\qquad +\,
\myint d^2\th d^2\bar{\th}\, \scr{Z}_M(u^\dagger, u) \left(
M^\dagger e^{v_D} M + \bar{M}^\dagger e^{-v_D} \bar{M} \right)
\\
&\qquad +\,
\left[ \myint d^2\th\, \left(
\sqrt{2} \bar{M} a_D(u) M + \La_S^2 m u \right) + \hc \right]
\\
&\qquad +\,
O(M^4) + \hbox{derivative\ terms}.
\eal\eeq
A term of the form \cite{AGLR}
\beq
\myint d^2\th\, w_D^\al \bar{D}^2 D_\al \ln \hat{\scr{Z}}_\Phi + \hc
\eeq
is absent by charge conjugation.
(A careful discussion of charge conjugation is given below.)
The monopole fields form a $SU(2)_R$ doublet
\beq
\scr{M}^j = \pmatrix{M \cr i\bar{M}^\dagger \cr}.
\eeq
Note that $M$ is in a doublet with $\bar{M}^\dagger$
(rather than $\bar{M}$) because the doublet must have well-defined
$U(1)$ gauge charge.
The factor of $i$ is required by $SU(2)_R$:
the coefficient of the term $\psi_{\bar{M}} a_D \psi_M$ is real,
while the coefficient of the terms $\psi_{\bar{M}} \psi M$ and
$\psi_{\bar{M}} \la M$ are both imaginary.

The Seiberg--Witten prepotential gives the exact
effective \Kahler potential
and gauge coupling with the monopoles integrated out.
Since the effective theory \Eq{mleff}
includes the monopole fields, we must
`integrate in' the monopoles, {\it i.e.}
invert the process of integrating out the monopoles.
In a general theory this is not unique, but in the present case
we need the effective \Kahler potential
and gauge coupling only in the $\scr{N} = 2$ limit,
where the result of integrating out the monopoles is exhausted by a
1-loop calculation.
We therefore have
\beq\eql{skintin}
\bal
s_D^{\vphantom{\rm (SW)}}(\mu) &= s_{D}^{\rm (SW)}
+ \frac{1}{8\pi^2} \left(
\ln\frac{a^{\vphantom{\dagger}}_D}{\mu}
+ 1 + c \right),
\\
k_D(\mu) &= k^{\rm (SW)}
+ \frac{a_D^\dagger a^{\vphantom{\dagger}}_D}{8\pi^2}
\left( \ln \frac{a_D^\dagger a^{\vphantom{\dagger}}_D}{\mu^2}
+ c \right),
\eal\eeq
where `SW' denotes the Seiberg--Witten solution with the monopoles
integrated out, and $c$ is a scheme-dependent constant.
(The analytic corrections to $s_D$ and $k$ are related
by $\scr{N} = 2$ SUSY.)
We will set $c = 0$ from now on.
Here $\mu$ is a renormalization scale that is required because the
effective theory containing the monopoles
has marginal interactions, and hence logarithmic
renormalization effects.
As a consistency check, we note that $s_D$ and the \Kahler metric
$(k_D)_{u^\dagger u}$ derived from \Eq{skintin}
are non-singular as $u \to 1$ ($a_D \to 0$).

In order to determine $\scr{Z}_M$ in \Eq{mleff}, we must discuss
the transformation of the monopole fields under the
anomalous $U(1)$ transformation \Eq{anomu1}.
Note that the anomalous
$U(1)$ is broken both explicitly (by anomalies) and
spontaneously (by $\avg{u} \ne 0$).
Furthermore, the monopole fields are not in any sense
simple functions of UV fields, so we must proceed carefully.
The most general transformation law allowed by holomorphy,
$U(1)$ gauge invariance, and dimensional analysis is
\beq
M \mapsto f(A, u, \bar{M} M / \La_S^2) \cdot M,
\quad
\bar{M} \mapsto \bar{f}(A, u, \bar{M} M / \La_S^2)
\cdot \bar{M},
\eeq
where $A$ is the anomalous $U(1)$ transformation parameter.
The explicit (anomalous) breaking of $U(1)_A$ is contained entirely
in the fact that $\La_S$ is not invariant.
The monopole term in the superpotential is therefore $U(1)_A$
invariant, which gives
\beq
\bar{f} f = e^{-A}.
\eeq
(Recall that $a_D \propto \La_S$, so $a_D$ has charge $+1$.)
Finally, charge conjugation (defined below) exchanges $M$ and
$\bar{M}$, and therefore implies $\bar{f} \equiv f$.
We conclude that the monopole and antimonopole transform linearly
under the anomalous $U(1)$ with the same charge $-\frac{1}{2}$.
[The details of the charge conjugation argument are as follows.
Define $C$ in the ultraviolet theory as
\beq
C:\ V \mapsto -V^T,
\quad
\Phi \mapsto +\Phi^T.
\eeq
This is a symmetry of the UV lagrangian, and
the positive sign for $\Phi$ is chosen so that $\avg\Phi \ne 0$ does
not break $C$.
(In a manifestly $\scr{N} = 2$ symmetric description, $C$ is therefore
an $R$ symmetry.)
The coupling spurions $\scr{Z}_\Phi$ and $S$ are clearly invariant
under $C$.
When the $SU(2)$ gauge group breaks to $U(1)$, the fields
in the effective theory transform as
\beq
C:\ v \mapsto -v,
\quad
u \mapsto u,
\eeq
where $v$ is the $U(1)$ gauge superfield.
This is obvious far from the origin where the theory is
weakly coupled, and cannot change in the strong-coupling
region because of continuity.
Because the monopole fields have opposite charge under the
`dual' $U(1)$ gauge group, they must transform as
\beq
C:\ M \leftrightarrow \pm \bar{M}
\eeq
under $C$, implying $\bar{f} \equiv \pm f$.
The negative sign is ruled out by the fact that $f = +1$ for
the identity transformation $A = 0$.]

Now that we know how the monopole fields transform under the anomalous
$U(1)$ transformation \Eq{anomu1}, we can fix $\scr{Z}_M$ in \Eq{mleff}
as a function of the UV couplings.
$\scr{Z}_M$ does not run by $\scr{N} = 2$ SUSY, so it must be an
RG-invariant function of $\hat{\scr{Z}}_\Phi$ and $\La$.
It cannot depend on $\La$ by dimensional analysis, and the anomalous
$U(1)$ tells us
\beq\eql{wavemono}
\scr{Z}_M = c \hat{\scr{Z}}_\Phi^{-1/2},
\eeq
where $c$ is a constant that is fixed by the $\scr{N} = 2$ limit:
\beq
\left.\scr{Z}_M \right|_{\th = \bar{\th} = 0} = 1.
\eeq
This result can also be obtained using the gauged non-anomalous $U(1)_R$
described in Section 3.
One has $R_\Phi = 0$ so that $\hat{\scr{Z}}_\Phi=e^{-2V_R/3}$.
On the other hand, by charge symmetry and $R$-invariance of the
low-energy theory, the monopole $R$ charges are $R_M = R_{\bar M} = 1$.
Therefore $\scr{Z}_M = e^{V_R/3}$ consistent with \Eq{wavemono}.

A simple but remarkable consequence of these results is that the
monopole soft mass does not run to all orders in perturbation theory
in the low-energy theory.
This is {\it a priori} surprising because the theory has no unbroken
SUSY and has marginal interactions.
The reason is simply that the wavefunction parameter of the monopoles
does not run in the $\scr{N} = 2$ limit.
The running of the soft masses is obtained by analytically continuing
the running in the SUSY limit into superspace \cite{AGLR}, and is therefore
controlled by the SUSY limit.

Straightforward calculation gives the potential to be
\beq\eql{monpot}
\bal
V &= V_0 + (2 |a_D|^2 - \sfrac{1}{2} m_{\phi 0}^2)
(|M|^2 + |\bar{M}|^2)
+ \frac{g_D^2}{2} (|M|^2 + |\bar{M}|^2)^2
\\
&\qquad -\frac{\sqrt{2} g_D^2 \La}{4\pi^2}
\left (\frac{\partial a_D}{\partial u}\right )^{-1}
\left[ \frac{4\pi^2 m_{\la 0}}{g_0^2} \bar{M} M
+ \frac{4 \pi^2 m_{\chi 0}}{g_0^2} (\bar{M} M)^\dagger \right] + \hc,
\eal\eeq
where $V_0$ is the potential given in \Eqs{theV} and \eq{thef}
with the replacement $k \to k_D$, and the running dual gauge coupling is
$1/g_D^2 \equiv [s_D]_0$ (see \Eq{skintin}).
Note that this potential is $SU(2)_R$ invariant, since
\beq\bal
\scr{M}^\dagger_j \scr{M}^j &= |M|^2 + |\bar{M}|^2,
\\
\scr{M}^\dagger_j \ep^{jk} (m_{\Psi 0})_{k\ell} \scr{M}^\ell
&= -i \left( m_{\la 0} \bar{M} M
+ m_{\chi 0} (\bar{M} M)^\dagger \right).
\eal\eeq

We now consider the energetics of the potential near the monopole points.
For this purpose, it is convenient to expand the potential in powers of
$u' = u - 1\simeq 2 i a_D$ and write
\beq\eql{bpot}
\bal
V &= g_D^2 \left|
\sqrt{2} \bar{M} M +\frac{2i \La}{g_0^2} (m_{\la 0}^* - m_{\chi 0})
\right|^2
+ \frac{g_D^2}{2} (|M|^2 - |\bar{M}|^2)^2
\\
&\qquad
+ \left[ -\sfrac{1}{2} m_{\phi 0}^2 + \sfrac{1}{2}|u'|^2\Lambda ^2 
+ O(u'^4) \right] (|M|^2 + |\bar{M}|^2)
\\
&\qquad
+ V_{\rm Naive}(u = 1) + \La^2 \left[ \left(
m_{\phi 0}^2 k_u(u = 1) + 8\pi^2 \De_0 + \frac{m_{B0}^2}{g_0^2} \right) u'
+ \hc \right]
\\
&\qquad
+ \scr{O}(m_{\rm soft}^2 u'^2),
\eal\eeq
where
\beq
V_{\rm Naive}(u = 1) = V_0(u = 1)
+ 4 g_D^2 \La^2 \left|
\frac{m_{\la 0} - m_{\chi 0}^*}{g_0^2} \right|^2
\eeq
is the potential with the monopole fields set to zero.
The above form of the potential allows us to easily understand
the origin of the cusp singularities in Figs.~2 and 3.
These arise if we set $M = \bar{M} = 0$ and evaluate the running
coupling $g_D^2(\mu)$ at a renormalization scale equal to
the supersymmetric monopole mass of order $|a_D|$.
Since $g_D^2(\mu) \sim 1/\ln\mu$ for $\mu\ll\La$, this gives a
logarithmic singularity as $a_D \to 0$.
This singularity is smoothed out when we minimize the full potential
because the monopole masses do not go to zero as $a_D \to 0$ in the
presence of soft SUSY breaking.
(The quantum corrections to the effective potential are well approximated
by evaluating the running coupling $g_D^2$ at a renormalization scale
of order the monopole VEV.)

We now turn to the monopole VEV's.
Assuming that $m_{\la 0} - m_{\chi 0}^*$ is nonzero and all soft
masses are of the same order,
the monopole VEV's are essentially determined by minimizing the first
two terms as long as $|u'|^2 \ll |m_{\la 0} - m_{\chi 0}^*| / \La$.
This gives
\beq\eql{monvev}
|\avg{M}|^2 \simeq |\avg{\bar{M}}|^2 \simeq {\sqrt 2}\La\,
\left |\frac{m_{\la 0}^* - m_{\chi 0}}{g_0^2}\right |.
\eeq
Note that this justifies dropping the $\scr{O}(M^4)$ terms in the
effective lagrangian \Eq{mleff}, since they contribute to the vacuum
energy at most $m_{\rm soft}^2 \avg{M}^4 \sim m_{\rm soft}^4$.%
\footnote{Higher order terms in the monopole fields are severely
constrained because the monopole fields are short $\scr{N} = 2$
multiplets.
However, it is interesting to note that we do not need the power
of $\scr{N} = 2$ SUSY to justify dropping these terms.}
The perturbation $m_{\la 0}^* - m_{\chi 0}$ is equivalent
to an $\scr{N} = 1$ superpotential mass, so the system is close to the
confining phase found in \Ref{SW}.

The monopole VEVs induce a positive mass-squared for $u'$ of order
$\La |m_{\la 0}^* - m_{\chi 0}|$.
This is larger than the $\scr{O}(m_{\rm soft}^2)$ contributions neglected
in \Eq{bpot}, so this stabilizes the modulus at $u' \sim m_{\rm soft}$.
Therefore the modulus is near the monopole points, and the approximations
made above are consistent.

We now consider the effect of the monopole VEVs on the vacuum energy.
This is important for determining whether there are first order
phase transitions between the monopole points and other local
minima on the moduli space. By the above qualitative discussion,
it is easy to conclude that the value of the potential at the minimum
near the monopole points
is $V_{\rm Naive}(u=1) + {\cal O}(\Lambda m_{\rm soft}^3)$.
To obtain this result, note
that the terms in the first  line of \Eq{bpot} respect ${\cal N} = 1$
SUSY, and therefore almost cancel at the minimum. Their contribution is
therefore only ${\cal O}(m_{\rm soft}^3)$ instead of 
${\cal O}(m_{\rm soft}^2)$.

This result shows that the vacuum energy near the monopole points
is well approximated by the value at the cusp singularities in the
\naive potential given in Figs.~1--4 that neglects the monopoles.
Therefore, we can use Figs.~1--4 to decide if the vacuum is in the
monopole (or dyon) region.%
\footnote{In Figs.~1--4 we plot $V / m_{\rm soft}^2$;
since $\partial^2 V / \partial u^2 = {\cal O}(m_{\rm soft})$ at
the monopole minimum, the second derivative is large in this
plot.}
%
At these points
there are always at least local minima with
\beq
\avg{M}^2, \avg{D}^2 \simeq
{\sqrt 2}\La \left| \frac{m_{\lambda 0} \mp m_{\chi 0}^*}{g_0^2} \right|
=  \frac{\sqrt{2} \La}{4\pi^2} \left( m_{\rm soft, 2}^2 \mp 2 m_{\rm soft, 3}^2
\right)^{1/2}.
\eeq
When $m_{\rm soft,1}^2$, $m_{\rm soft,5}^2$, $m_{\rm soft,6}^2$
are sufficiently smaller than the other soft terms, we have that
for $m_{\rm soft, 3}^2 < 0$ the global minimum is
at the monopole point, while for $m_{\rm soft, 3}^2 > 0$ it is at the dyon
point. This can be seen from Figs.~2 and 3. Notice also that
in the limiting cases
$m_{\rm soft, 2}^2 = \pm 2 m_{\rm soft, 3}^2$ the local
monopole (dyon) VEV disappears \cite{ADKM}.
On the other hand, when $m_{\rm soft, 1}^2$ is sufficiently larger
than the other soft terms the local minimum at the origin $u = 0$
is the global minimum.
Indeed,
when $m_{\rm soft,1}^2$ dominates, the monopole vevs are even less 
important to the vacuum energetics.
As can be inferred from \Eq{bpot}, 
they can decrease the energy only for $|u'|^2<m_{\phi 0}^2/\Lambda^2$
and only by an amount $\Delta V\sim -m_{\phi 0}^4/g_D^2$.


\section{Conclusions}
We have considered the most general soft SUSY breaking of $\scr{N} = 1$
and $\scr{N} = 2$ theories, including non-holomorphic perturbations.
Using the method of \Ref{AR} we are able to obtain exact results when
the soft masses are small compared to the scale of non-perturbative
physics ($m_{\rm soft} \ll \La$) because SUSY relates soft mass terms
to background gauge fields.
We gave a new formulation of this correspondence in terms of a
non-anomalous gauged $U(1)_R$ symmetry in a supergravity background.
We also applied this formalism to several cases of interest:
$\scr{N} = 1$ theories deformed moduli spaces and conformal fixed points,
and $\scr{N} = 2$ super-Yang--Mills.

Our results show that in many cases, the theory
for $m_{\rm soft} \ll \La$ is in a different phase than the
$m_{\rm soft} \to \infty$ limit.
For example, in the $\scr{N} = 2$ $SU(2)$ super-Yang--Mills theory,
adding a small soft scalar mass drives the theory to a free Coulomb
phase, while we believe that the $m_{\rm soft} \to \infty$ theory
is in a confining phase.
This means that there are necessarily phase transitions as a
function of the soft masses at $m_{\rm soft} \sim \La$.
For example, this is important for non-perturbative studies of these
models on the lattice, where supersymmetry presumably has to be
imposed by tuning lattice parameters.
Clearly, the road to understanding the relationship between
supersymmetric and non-supersymmetric gauge theories remains a long
one, but we hope that the steps taken in this paper will prove
useful.

\section*{Acknowledgments}
We thank Nima Arkani-Hamed, Ann Nelson,  Matt Strassler, and Scott Thomas  
for discussions.
M.A.L. was supported by the National
Science Foundation under grant PHY-98-02551,
and by the Alfred P. Sloan Foundation.


\end{document}